\documentclass[sn-mathphys,Numbered]{sn-jnl}
\usepackage{appendix}
\usepackage{geometry}
\usepackage{import}
\usepackage{float}
\usepackage{amsmath,amsfonts,amssymb,amsbsy,amsthm}
\usepackage{mathtools}
\usepackage{hyperref}
\usepackage{subfig}
\usepackage{multirow}
\usepackage{multicol}
\usepackage{etoolbox}
\usepackage{algorithm}
\usepackage[noend]{algpseudocode}
\usepackage[dvipsnames]{xcolor}
\usepackage{paralist}
\allowdisplaybreaks
\usepackage{booktabs}
\usepackage[group-separator={,}]{siunitx}
\usepackage{textcomp}
\usepackage{mathrsfs}
\usepackage{ifpdf}
\usepackage{lmodern}
\usepackage[LY1]{fontenc}

\usepackage{xcolor}%
\usepackage{manyfoot}%
\usepackage{booktabs}%
\usepackage{algorithmicx}%
\usepackage{algpseudocode}%
\usepackage{listings}%

\usepackage{graphicx}

\usepackage{caption}

\theoremstyle{thmstyleone}%
\theoremstyle{thmstyletwo}%

\theoremstyle{thmstylethree}%

\newcommand{\SGU}[1]{{{[SU]}}}

\hypersetup{hidelinks=true}

\definecolor{arsenic}{rgb}{0.23, 0.27, 0.29}
\definecolor{charcoal}{rgb}{0.21, 0.27, 0.31}
\definecolor{hanblue}{rgb}{0.27, 0.42, 0.81}
\definecolor{blue-ncs}{rgb}{0.0, 0.53, 0.74}
\definecolor{awesome}{rgb}{1.0, 0.13,0.32}
\definecolor{darkgreen}{rgb}{0, .4,0}

\newcommand{\vect}[1]{{\boldsymbol{\mathbf{#1}}}}

\begin{document}

\title[A GPU-accelerated simulator for the DEM analysis of granular systems composed of clump-shaped elements]{A GPU-accelerated simulator for the DEM analysis of granular systems composed of clump-shaped elements}

\author[1]{\fnm{Ruochun} \sur{Zhang}}\email{rzhang294@wisc.edu}

\author[1]{\fnm{Colin} \sur{Vanden Heuvel}}\email{colin.vandenheuvel@wisc.edu}

\author[2]{\fnm{Alexander} \sur{Schepelmann}}\email{alexander.schepelmann@nasa.gov}

\author[3]{\fnm{Arno} \sur{Rogg}}\email{arno.rogg@nasa.gov}

\author[4]{\fnm{Dimitrios} \sur{Apostolopoulos}}\email{da1v@protoinnovations.com}

\author[4]{\fnm{Samuel} \sur{Chandler}}\email{samxchandler@protoinnovations.com}

\author[1]{\fnm{Radu} \sur{Serban}}\email{serban@wisc.edu}

\author*[1]{\fnm{Dan} \sur{Negrut}}\email{negrut@wisc.edu}

\affil[1]{\orgdiv{Department of Mechanical Engineering}, \orgname{University of Wisconsin-Madison}, \orgaddress{\street{1513 Engineering Dr}, \city{Madison}, \postcode{53706}, \state{WI}, \country{USA}}}

\affil[2]{\orgdiv{Mechanisms and Tribology Branch}, \orgname{NASA Glenn Research Center, 23-217}, \orgaddress{\street{21000 Brookpark Road}, \city{Cleveland}, \postcode{44135}, \state{OH}, \country{USA}}}

\affil[3]{\orgdiv{VIPER Robotic Systems}, \orgname{KBRwyle, NASA Ames Research Center}, \orgaddress{\city{Moffett Field}, \postcode{94035}, \state{CA, MS 269-3}, \country{USA}}}

\affil[4]{\orgdiv{Terrain Manipulation and Wheels}, \orgname{ProtoInnovations}, \orgaddress{\street{100 43rd St}, \city{Pittsburgh}, \postcode{15201}, \state{PA}, \country{USA}}}


\abstract{
We discuss the use of the Discrete Element Method (DEM) to simulate the dynamics of granular systems made up of elements with nontrivial geometries. The DEM simulator is GPU accelerated and can handle elements whose shape is defined as the union with overlap of diverse sets of spheres with user-specified radii. The simulator can also handle complex materials since each sphere in an element can have its own Young's modulus $E$, Poisson ratio $\nu$, friction coefficient $\mu$, and coefficient of restitution CoR. To demonstrate the simulator, we produce a ``digital simulant'' (DS), a replica of the GRC-1 lunar simulant. The DS follows an element size distribution similar but not identical to that of GRC-1. We validate the predictive attributes of the simulator via several numerical experiments: repose angle, cone penetration,  drawbar pull, and rover incline-climbing tests. Subsequently, we carry out a sensitivity analysis to gauge how the slope vs. slip curves change when the element shape, element size, and friction coefficient change. The paper concludes with a VIPER rover simulation that confirms a recently proposed granular scaling law. The simulation involves more than 11 million elements composed of more than 34 million spheres of different radii. The simulator works in the Chrono framework and utilizes two GPUs concurrently. The GPU code for the simulator and all numerical experiments discussed are open-source and available on GitHub for reproducibility studies and unfettered use and distribution.
}


\pacs[Article Highlights]{
\newline
1. An open-source and publicly available dual-GPU DEM simulator is discussed, which handles elements composed of sets of clumped spheres of potentially different radii and material properties.
\newline
2. A variety of tests concerning soil response and wheel mobility performance confirm the predictive attribute of the simulator, and its ability to scale to large granular problems.
\newline
3. A digital replica of the GRC-1 lunar simulant is developed and simulated using this solver. The DEM solution accurately captures the terramechanical properties of GRC-1.
}

\pacs[Statements and Declarations]{This work has been partially supported by NSF projects OAC2209791 and CISE1835674, US Army Research Office project W911NF1910431, and NASA Small Business Technology Transfer (STTR) Sequential Contract \#80NSSC20C0252.}

\maketitle

\newpage

\section{Introduction} \label{sec:intro}

The Discrete Element Method (DEM) is a computational approach used to predict the mechanical behavior of granular materials \cite{cundall1979discrete}. DEM keeps track of the motion of each individual element and models the elements' mutual interactions in a fully resolved fashion. DEM has been extended over time and has become a widely-used approach for predicting the dynamics of large granular systems~\cite{poschelDEM-textbook2005}, from~mixing~\cite{lemieux2008large}, and particulate flows~\cite{apostolou2008discrete}, to landslides and other geomechanics phenomena \cite{tang2009tsaoling, salciarini2010discrete,OSullivan2011}, and astrophysical processes~\cite{sanchez2011simulating}. DEM has been used to simulate soil behavior \cite{frederikOleDEM2022}, the tire-terrain interaction \cite{antonioVehicleTireGranMatSim2017}, and rover mobility in extraterrestrial conditions \cite{iagnema2015}. 

DEM simulations are generally slow due to two main reasons. First, the elements are often stiff and sometimes very small. This forces the time integrator to take very small time steps to maintain numerical stability. Secondly, collision detection is compute-intensive. To reduce simulation times, DEM has been accelerated via CPU-based parallel computing using OpenMP \cite{openMP}, see, for instance, \cite{amritkar2014efficient,knuth12}; the Message Passing Interface (MPI) standard \cite{mpi-3.0}, for clusters with distributed memory~\cite{yan2018comprehensive}; and hybrid MPI--OpenMP parallelism~\cite{checkaraou2018hybrid,liggghts2013,lammpsWebSite,raduNicDanGVSETS2018}. An alternative computing architecture for parallel computing is provided by the Graphics Processing Unit (GPU), which has gained wide adoption in DEM, e.g., \cite{xu2011quasi, govender-BlazeDEMGPU2016,ganDEM-GPU2016,he-powderGPU2018}. Whether using CPU or GPU computing, the number of DEM elements used in experiments reported in the literature has been relatively small -- in the vicinity of $10^3$ to $10^5$ elements~\cite{iwashitaRolling1998,direnzo2004525,dacruz2005,bazant2006,Emden2008,wasfy2014coupled,lommen2014,Utili2015,Potticary2015,michael2015,Ciantia2016,zheng2017,parteli2016,kivugo2017,calvetti2016,he-powderGPU2018}. For comparison, in~one cubic meter of sand, there are on the order of two billion elements and simulations on this scale have rarely been conducted \cite{japanDEMlarge2018}. 

\begin{figure}[htp!]
	\centering
	\captionsetup{justification=centering}
	\includegraphics[width=.35\linewidth]{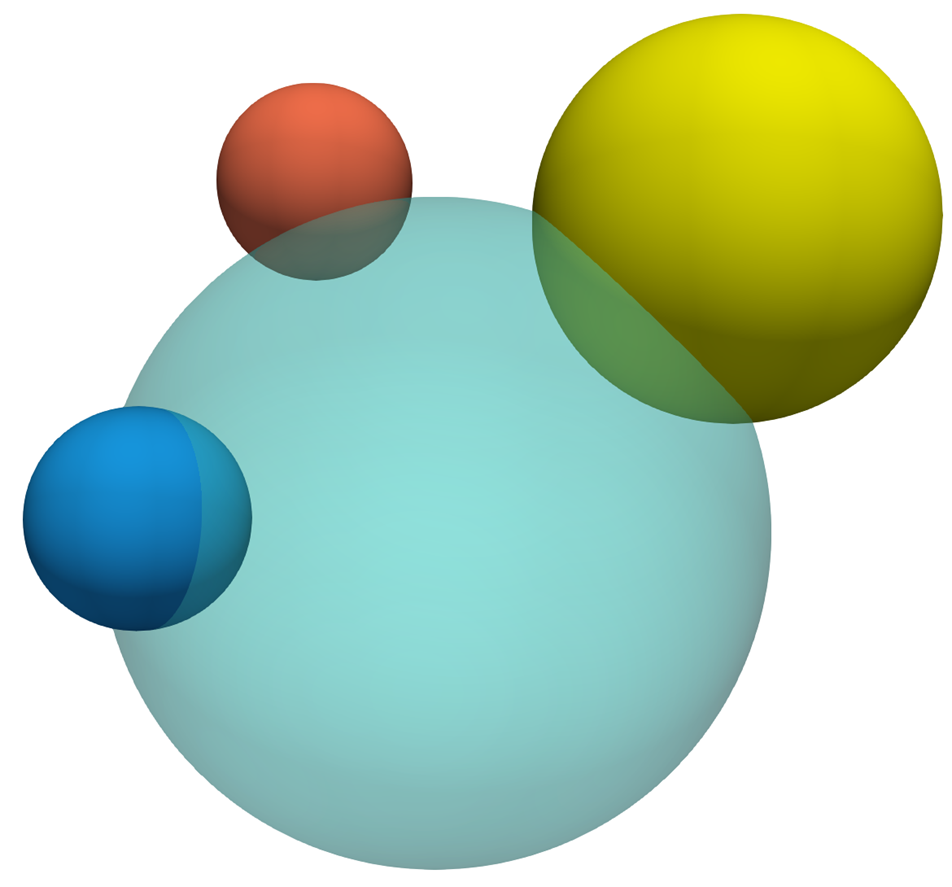}
	\caption{In the proposed approach, each element in the DEM simulation is a clump of spheres. The different colors emphasize the fact that each sphere in the clump has its own physical attributes, i.e., $E$, $\nu$, $\mu$, and CoR.} \label{fig:clump}
\end{figure}

In addition to utilizing parallel computing to control simulation times, practitioners typically rely on simple element geometries to avoid expensive collision detection calculations. By a large margin, the geometries employed in DEM are thus spheres, in many cases of equal radii. By exclusively dealing with spheres, the collision detection workload is significantly reduced. The software implementation process is also simpler, since handling nontrivial geometries is difficult \cite{Ericson05}. However, in many applications, as will be demonstrated in this contribution, one cannot fall back on simple element geometry without compromising the system response \cite{favierClumpsSpheres1999,particleShapeCleary2011,particleShapeKiangi2013,unionOf2Spheres2013,dem-cfdMonashAustralia2016,andradeDEM2018,marteauExperimental2021}. Yet in studies in which nontrivial element shapes are brought to bear, the number of simulated DEM elements reported in the literature drops significantly. Commercial solutions to handle non-trivial shapes exist, e.g., \cite{edemAltair}, but they are closed-source, expensive, and limited in their ability to handle systems with large element counts. In this context, the solution proposed herein embraces the following compromise: the number of DE can be large (into tens of millions) and while the shape of the elements is nontrivial, it is nonetheless assumed that each element is obtained as the union of a user-specified set of spheres of user-specified radii and material properties, see Fig.~\ref{fig:clump}. The idea of using multiple spheres to generate nontrivial geometries is not new, see, for instance, the discussion in \cite{dem-cfdMonashAustralia2016}, where the authors called the technique the ``multi-sphere method'', a member of the class of composite approaches in which elements' geometries are composed as the union of simple geometric primitives. However, previous attempts to use this multi-sphere method were limited to small systems. In \cite{particleShapeKiangi2013,multiSphereDEM-maione2015,GUO2022293}, the DE systems have a number of elements in the low thousands; in \cite{favierClumpsSpheres1999}, the authors work with hundreds of elements. The approach outlined here builds on two previous contributions: a fast way of doing collision detection, which follows in the step with the approach reported in \cite{hammadTobyDan2012}; and an aggressive use of shared memory along with a new way of representing state information using new data types, see \cite{conlainNicDan2020,chronoGranular2021}. The simulator is demonstrated in conjunction with a terramechanics problem, in which the terrain is modeled as a collection of elements each composed of clumped spheres. Given that the terrain model is inspired by the GRC-1 lunar simulant, it is called ``digital simulant'' (DS). 


The manuscript is organized as follows. 
Section~\ref{sec:soil_props} details the digital simulant. 
Section~\ref{sec:DEMModel} briefly introduces the DEM model and the DEM simulator.  
Section~\ref{sec:soil_tests} reports on validation studies, and touches on angle of repose, cone penetration, and drawbar pull simulations using the DS. The validation is done against experimental results that involved the GRC-1 simulant.
In Sec.~\ref{sec:rover_tests}, the simulator is used for extraterrestrial rover single-wheel and full-vehicle tests.
Section~\ref{sec:summary} contains conclusions and outlines directions of future work.  

\section{Simulator overview: force model and implementation aspects} \label{sec:DEMModel}

\subsection{Contact force model and clump representation} \label{sec:model}

The default force model in Chrono's DEM simulator draws on Hertzian contact \cite{hertz1882} and the Mindlin friction model \cite{mindlin53}. For an extended discussion see \cite{luningFricModel2021}. When bodies $i$ and $j$ are in contact, the mutual normal force $\vect{F}_n$ is described via a spring--dashpot mechanism. The tangential friction force $\vect{F}_t$ is evaluated based on material properties and local micro-deformations, and capped to satisfy the Coulomb condition through friction coefficient $\mu$.  Specifically,
\begin{subequations}
	\begin{align}
		&\vect{F}_n = f(\bar{R},\delta_n) (k_n \vect{u}_n - \gamma_n \bar{m} \vect{v}_n),  \label{eq:normal_tangential_force_model1} \\
		&\vect{F}_t = f(\bar{R},\delta_n)(-k_t \vect{u}_t -  \gamma_t \bar{m} \vect{v}_t),  \quad	\|\vect{F}_t\| \leq \mu \|\vect{F}_n\| \; , \label{eq:normal_tangential_force_model2} \\
		&f(\bar{R},\delta_n) = \sqrt{\bar{R}\delta_n}, \\
		&\bar{R} = R_i R_j/(R_i + R_j),\\
		&\bar{m} = m_i m_j/(m_i + m_j),
	\end{align}
\end{subequations}
where the stiffness and damping coefficients $k_n$, $k_t$, $\gamma_n$, and $\gamma_t$ are derived from material properties, i.e., Young's modulus $E$, Poisson ratio $\nu$, and the coefficient of restitution $\text{CoR}$ \cite{jonJCND2015}. The $ \bar{m}$ and $\bar{R}$ terms represent the effective mass and effective radius of curvature for this contact, respectively. 
The fundamental assumption is that geometries are allowed to have small penetration $\delta_n$ at the point of contact.
The normal penetration vector is defined as $\vect{u}_n = \delta_n \vect{n}$. The relative velocity at the contact point $\vect{v}_{rel} = \vect{v}_n + \vect{v}_t$ are evaluated as
\begin{subequations}
	\label{subeq:kinematicsContactInfo}
	\begin{align}
	\vect{v}_{rel}& = \vect{v}_j + \vect{\omega}_j \times \vect{r}_j - \vect{v}_i - \vect{\omega}_i \times \vect{r}_i, \\
	\vect{v}_n &= \left(\vect{v}_{rel} \cdot \vect{n} \right) \vect{n}, \\
	\vect{v}_t &= \vect{v}_{rel} - \vect{v}_n,
	\end{align}
where $\vect{v}_{i}$, $\vect{\omega}_{i}$ and $\vect{v}_{j}$, $\vect{\omega}_{j}$ are the velocity at the center of mass and the angular velocity of bodies $i$ and $j$, respectively. The position vectors $\vect{r}_{i}$ and $\vect{r}_j$ point from the center of mass of bodies $i$ and $j$  to the mutual contact point. 
The friction force $\vect{F}_t$ depends on the tangential micro-displacement history $\vect{u}_t$ \cite{mindlin53,jonJCND2015}, the latter updated incrementally at each time step throughout the life of a contact event based on the tangential velocity $\vect{v}_t$. Let $\vect{u}^\prime_t$ be the updated tangential micro-displacement, then
	\begin{align} \label{subsubeq:updateUt}
	\vect{u}^\prime &=\vect{u}_t  + h\vect{v}_t ,\\
	\vect{u}^\prime_t &= \vect{u}^\prime - (\vect{u}^\prime \cdot \vect{n})\vect{n},
	\end{align}
where $h$ is the time step size. The strategy adopted to update $\vect{u}^\prime_t$ is borrowed from \cite{jonJCND2015}.
After the update, we may need to clamp the updated tangential micro-displacement $\vect{u}^\prime_t$ to get the final $\vect{u}_t $ for the next time step in order to satisfy the capping condition $\|\vect{F}_t\| \leq \mu \|\vect{F}_n\|$:
	\begin{align} \label{subsubeq:cappingOfut}
	\vect{u}_t = \begin{cases} \vect{u}^\prime_t & \text{if } \|\vect{F}_t\| \leq \mu \|\vect{F}_n\|, \\
	 		 \frac{\mu\|\vect{F}_n\|}{k_t} \frac{\vect{u}^\prime_t}{\|\vect{u}^\prime_t\|} & \text{otherwise.} \end{cases}
	\end{align}
\end{subequations}

\begin{figure}[htp!]
	\centering
	\captionsetup{justification=centering}
	\includegraphics[width=.85\linewidth]{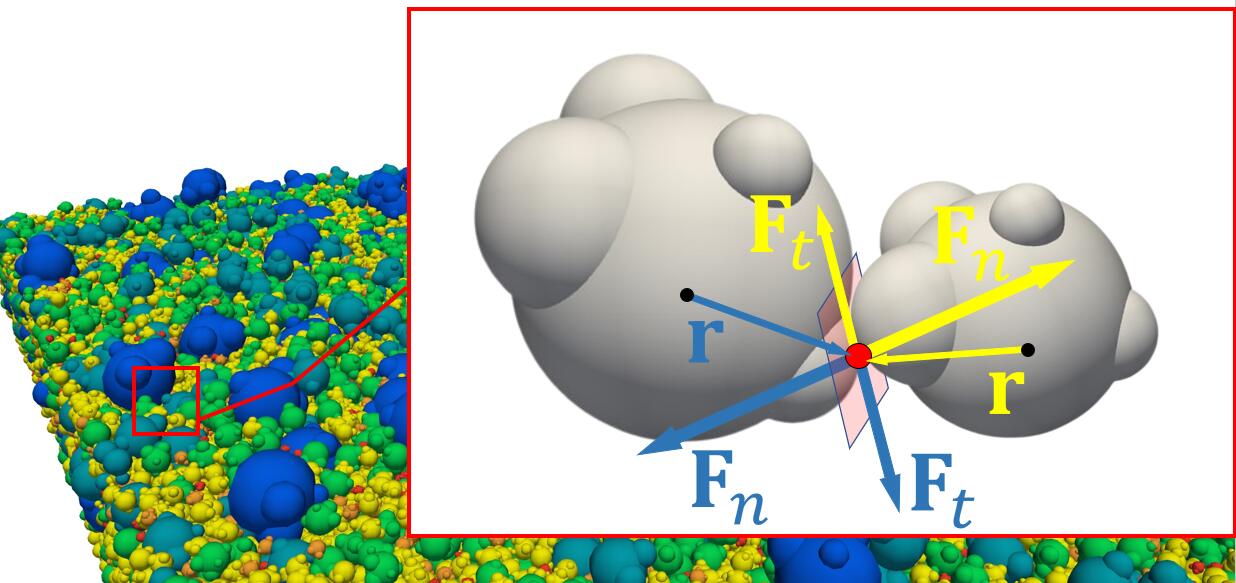}
	\caption{The normal and tangential contact forces between clumps are calculated based on the penetration and displacement history of involved sphere components. An element (clump) has a mass and mass moment of inertia; however, each sphere in the clump has its own $E$, $\nu$, $\mu$, and CoR.} \label{fig:force}
\end{figure}

The equations of motion for element $i$ assume the expression
\begin{subequations}
	\begin{align}
		m_i \frac{d \vect{v}_i}{d t} &= m_i \vect{g} + \sum_{j=1}^{nc} \left(\vect{F}_n + \vect{F}_t \right), \label{eq:newton_law_1} \\
		I_i \frac{d \vect{\omega}_i}{dt} &= \sum_{j=1}^{nc} \left( \vect{r}_j \times \vect{F}_t  \right). \label{eq:newton_law_2}
	\end{align}	
\end{subequations}
Chrono's DEM simulator supports complex element geometry through clumps using an approach borrowed from \cite{clumpSpheresPrice2007}. A clump is a collection of potentially overlapping spheres representing an element shape, see Fig.~\ref{fig:force}. Herein, the word ``clump'' is used interchangeably with ``element'' when referring to complex shape DEM elements. One salient attribute of the model is that the $E$, $\nu$, $\mu$, and CoR parameters are associated with a sphere in an element/clump, and not with the element. In numerous instances, the inclusion of this extensive set of $E$, $\nu$, $\mu$, and CoR parameters for the spheres within the clump may be deemed excessive. However, it remains a viable choice for those seeking to implement a sophisticated discrete element system.

\subsection{DEM simulator overview}

The current DEM simulator, whose design and implementation details are discussed at length in an upcoming contribution, is an upgrade to Chrono's previous DEM simulator \cite{chronoGranular2021}. Although fast \cite{dem-pbrIdaho2023}, the latter only handled monodisperse granular systems \cite{conlainBillion2019}. For a detailed discussion of the current DEM simulator, DEM-Engine, the reader is referred to \cite{RuochunDEMERepo}. Herein, the focus is on demonstrating Chrono's DEM simulator rather than discussing the design decisions that shaped its implementation. Note that the DEM simulator can run in standalone mode -- it is built and can run separately from Chrono. When it runs in co-simulation mode with Chrono, the latter runs the dynamics of the machine system, while the DEM simulation handles the granular dynamics. This opens the door to simulating complex mechanical systems, e.g., rovers, construction equipment, interacting with granular materials -- see numerical tests in Sec.~\ref{sec:full_rover}.

If the user prefers a different contact force model, he/she can provide a C++ callback customized function that overwrites the default one. Typically, the updating of the active contacts set and the DEM force calculation are done sequentially and at each time step. DEM-Engine embraces a different strategy, which uses two distinct and parallel computational threads to update the active contacts set (done by the ``kinematics thread'') and the integration of the equations of motion (done by the ``dynamics thread''), respectively. The user has the option to update the collision pairs at each time step, or can carry out several time steps before updating the contact pairs. In the latter case, the active contacts set is kept the same for several time steps. Regardless of whether the active contacts set is re-evaluated or not at each time step, each active contact is employed \textit{at each time step} to reassess the contact deformation $\delta_n$ and the ancillary information defined in Eqs.~\ref{subeq:kinematicsContactInfo}. This active contacts set delayed re-evaluation strategy was implemented since it became apparent that identifying the active set represents a significant computational bottleneck. To avoid missing mutual contacts that might crop up between the moments the active contacts set is updated, for the task of updating the active contact set only, we artificially enlarge all contact geometries in the DEM system to preemptively detect potential contact pairs that might emerge in the near future. 

At \textit{each} time step, the dynamics thread uses the most recent active contacts set to update the penetration $\delta_n$ and ancillary quantities (see Eqs.~\ref{subeq:kinematicsContactInfo}) needed to evaluate the set of frictional contact forces at work in the system. Note that by artificially enlarging the elements' geometries, the active contacts set will contain false positives and the dynamics thread will spend extra time calculating $\delta_n$ for a contact only to notice that the gap between the bodies is larger than zero in which case the frictional contact force is set to zero. The thickness of the added safety margin, which is dictated by the maximum clump velocity (bounded and known to the solver) and the time step size (typically small in large-scale DEM simulations), is a known value, e.g., for millimeter-sized granular material, it assumes values of the order of tens of microns. Overall, the overhead caused by the false-positive contacts does not offset the benefit of deferred updates of the active contacts set.
This is because the kinematics and dynamics threads work concurrently and the computational effort dispensed to update of the active contacts set by the kinematics thread happens ``in the shadow'' of the work done by the dynamics thread. Note that one still has the option to fall back on the traditional way of carrying out DEM simulation, in which the active contacts set is updated at each time step before computing the frictional contact forces and advancing the state of the system.


In terms of software implementation, the simulator is set up to use two NVIDIA GPUs -- one by the kinematics thread, one by the dynamics thread. It is more efficient to allow each thread to run on a dedicated GPU, yet if only one GPU is available, the simulator makes the two threads share the only GPU available via CUDA streams \cite{nvidiaCUDA}. As an alternative to using streams, starting with the NVIDIA Ampere architecture, one has the option to partition one GPU into several via NVIDIA's Multi-Instance GPU feature \cite{migNVIDIA2023}.

\begin{figure}[htp!]
	\centering
	\captionsetup{justification=centering}
	\includegraphics[width=.9\linewidth]{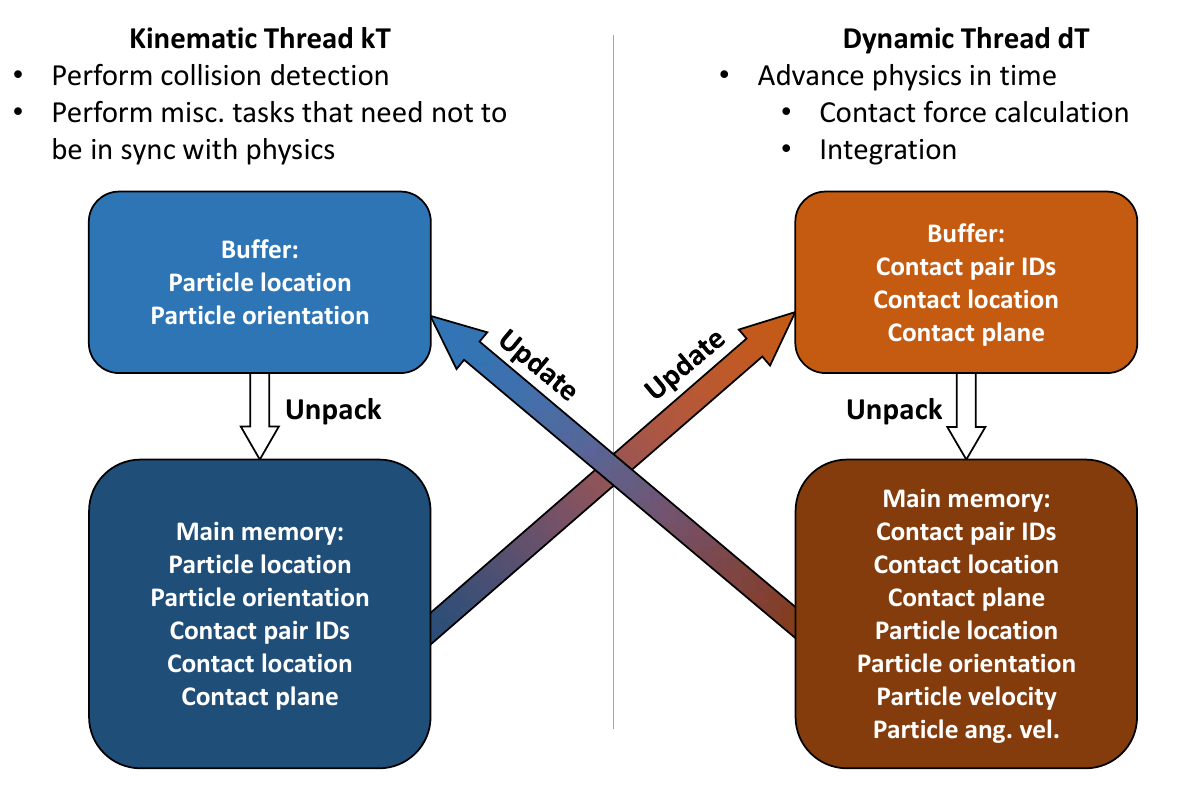}
	\caption{The collaboration pattern of the kinematic and dynamics thread. They each run on a dedicated GPU. Note if two GPUs are in use, the two buffers are both allocated physically on the GPU mapped to the dynamics thread.} \label{fig:kTdT}
\end{figure}

The collaboration pattern of the kinematics and dynamics threads is summarized in Figure~\ref{fig:kTdT}. After being produced by the kinematics thread, the active contacts set and associated bookkeeping information are written into a memory buffer to be read by the dynamics thread. The dynamics thread in turn updates the kinematics thread with new clump positions and velocities. Although logically there are two memory buffers, they are both allocated physically on the GPU mapped to the dynamics thread. This allows the dynamics thread to spend minimum time reading/writing data, thus speeding up the computation.

\section{Digital simulant used} 
\label{sec:soil_props}
The simulated granular material used in this manuscript, subsequently called digital simulant (DS), was inspired by the GRC-1 simulant~\cite{ORAVEC2010361}, developed as a lunar soil replica for Earth testing of rovers and similar implements. The GRC-1 simulant contains cohesionless and frictional particles that have complex shapes and span a range of sizes. The fact that some GRC-1 particle sizes are micron-scale posed insurmountable challenges to the simulator. Despite using a customized way of storing contact information that goes beyond what the IEEE double precision data type can offer in C/C++, see \cite{RuochunDEMERepo}, working with the actual particle size proved intractable owing to the broad spectrum of sizes -- from microns to millimeters. As a compromise, the digital simulant maintained the statistical size \textit{distribution}, yet uniformly increased by a factor of 20 the actual particle sizes encountered in GRC-1. In other words, the DS size distribution was identical to GRC-1, but the DS sizes were shifted to larger values. The value 20 mention above has no particular meaning other than being the smallest values that we had to scale by so that the DEM simulation became acceptable, both in terms of results accuracy and time to completion. The simulation experiments of Sec.~\ref{sec:sen_size} confirm that this ``scaling up'' of the particle sizes does not significantly compromise the accuracy of the results obtained through simulation. 

\begin{figure}[htp!]
\centering
\begin{minipage}{.45\textwidth}
	\centering
	\captionsetup{justification=centering}
	\includegraphics[width=\linewidth]{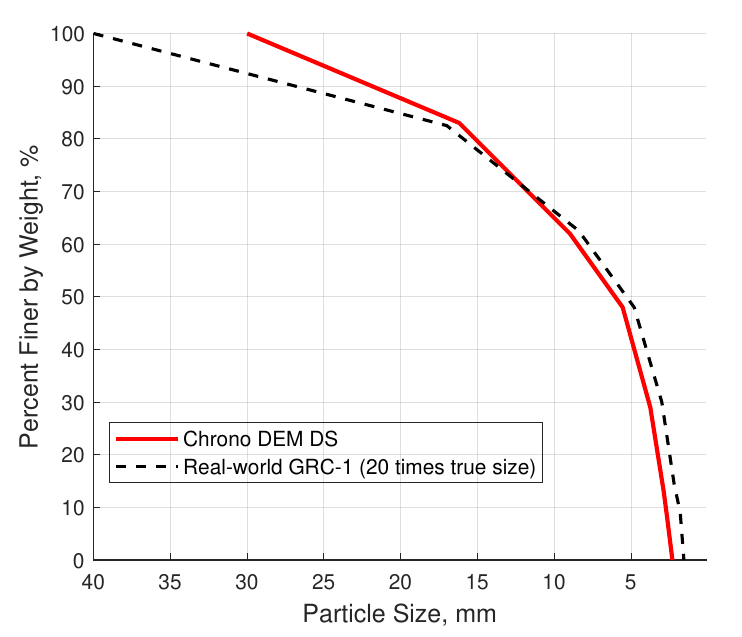}
	\caption{The size distribution of the DS, plotted against a scaled GRC-1 simulant size distribution.}\label{fig:GRCDS1}
\end{minipage}%
\hspace{.1cm}
\begin{minipage}{.45\textwidth}
	\centering
	\captionsetup{justification=centering}
	\includegraphics[width=.85\linewidth]{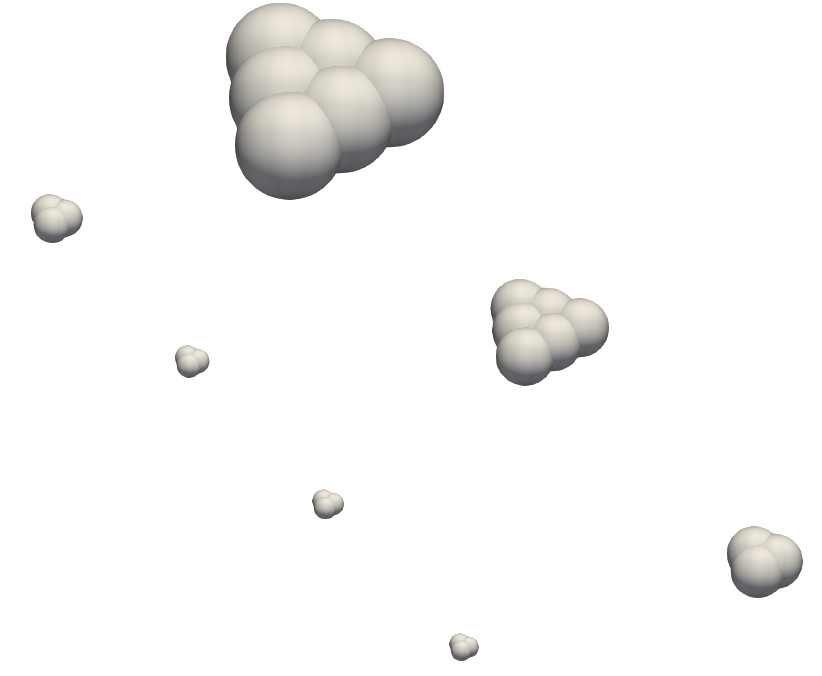}
	\caption{The seven clump shapes that show up in the DS.} \label{fig:GRCDS2}
\end{minipage}
\end{figure}

The DS consists of seven DEM element types, each with a specific size and percentage of the total weight, see Table~\ref{tab:GRCDS}. The size distribution is plotted in Fig.~\ref{fig:GRCDS1}. Therein, although the size distribution is not continuous, it is shown with seven line segments, each corresponding to the weight percentage contributed by each element type. Figure~\ref{fig:GRCDS2} illustrates the DS grain shapes. Out of the seven distinct grain types, the two larger ones are made of six overlapping spheres and have a flat triangular shape. The five smaller types are each made of three component spheres. All of them have a $120^\circ$ rotation symmetry. Having a triangular aspect, the particle size reported in Table~\ref{tab:GRCDS} is measured as the diameter of the bounding sphere. The radii of the elements' component spheres and the material properties used in the numerical tests throughout the paper are also summarized in Table~\ref{tab:GRCDS}.

\begin{table}
	\centering
		\begin{tabular}{*8l}    \toprule
			\emph{Type} & 1 & 2 & 3 & 4 & 5 & 6 & 7  \\\midrule
			Size (\SI{}{mm}) & 21  & 11.4  & 6.6  & 4.5 & 3 & 2.75 & 2.5 \\ 
			Component radius (\SI{}{mm}) &  3.6 & 1.95  & 1.81 & 1.24 & 0.82 & 0.75 & 0.7 \\ 
			\%, by weight & 17 & 21 & 14 & 19 & 16 & 5 & 8 \\ \bottomrule
			\hline
		\end{tabular}
	\caption{Weight distribution, percent-wise, by particle size. For all particle types, $E={\SI{e9}{N/m^2}}$, $\nu=0.3$, $\mu_s = 0.4$, and $\text{CoR}=0.5$.}
	\label{tab:GRCDS}
\end{table}

\begin{figure}[htp!]
	\centering
	\captionsetup{justification=centering}
	\subfloat[The single-wheel test on a $25^\circ$ incline, with DS.\label{fig:GRCat25}]
	{
		\includegraphics[width=0.45\linewidth]{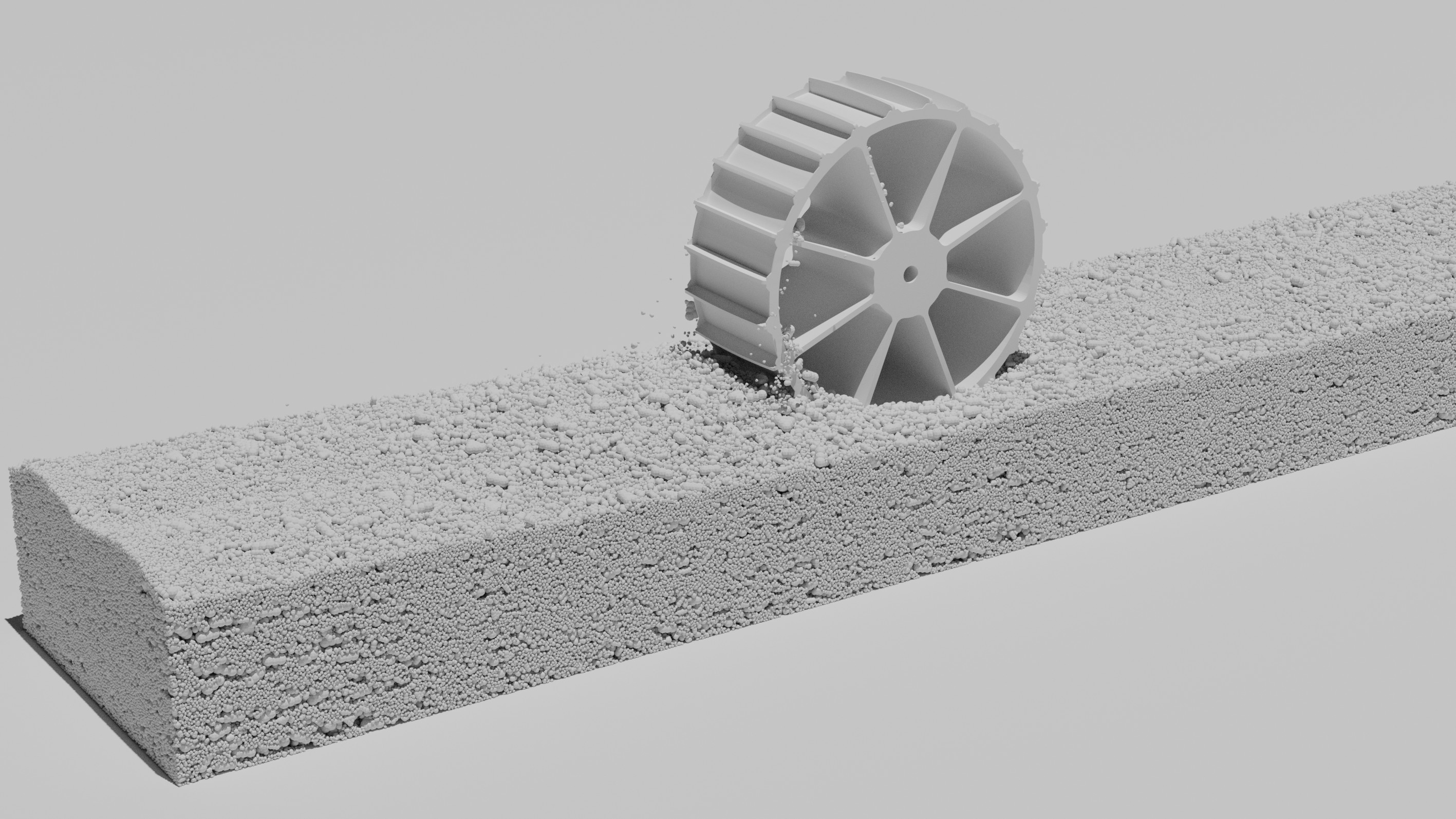}
	}
	\hspace{.1cm}
	\subfloat[The single-wheel test on a $25^\circ$ incline with spherical DEM elements. Snapshot suggests that monodisperse granular material support for a single wheel test is qualitatively different than what behavior observed for DS testing.\label{fig:SPHat25}]
	{
		\includegraphics[width=0.482\linewidth]{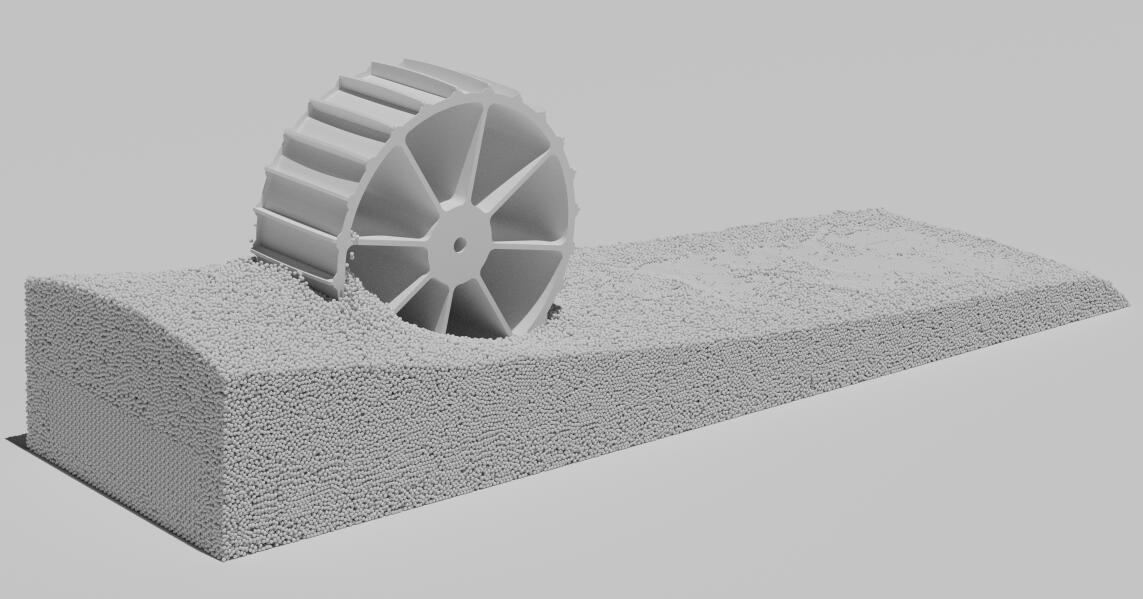}
	}
	\caption{The renderings show that, on the same $25^\circ$ incline, the single-wheel simulation on spherical DEM elements leads to much more terrain shearing compared to simulation on DS.} \label{fig:COMPat25}
\end{figure}

The nontrivial shape of the elements turned out to play an important role in the dynamics of the granular material. Compared to monodisperse granular material, which is commonly used in DEM simulations, the DS allowed for geometric locking, thus shaping the granular material's behavior, especially in shear-dominated experiments. Qualitatively, this observation is backed by the early results reported in Figs.~\ref{fig:GRCat25}~and \ref{fig:SPHat25}, wherein the same single-wheel drawbar pull experiment was carried out once using the DS and then monodisperse granular material.

The simulations shown in the following sections consist of millions to tens of millions of DEM elements, a scale that is manageable on one or two modern GPU devices. If the true GRC-1 grain sizes were used, the number of elements would increase by close to four orders of magnitude, which places an insurmountable barrier to simulation owing to memory constraints, as well as arithmetic precision limits as some deformations $\delta_n$ become prohibitively small. Note that the smallest sphere component in a DS clump had a radius of approximately \SI{700}{\micro m}.

\section{Validation studies} \label{sec:soil_tests}
The validation and refinement of a DEM simulator is an arduous and lengthy process that spans an extended period. Herein, we report on three experiments that were used for the early validation of the new DEM simulator in Chrono. The results in this section represent a confidence-building step before using the DEM simulator to conduct additional studies for which there is limited or no experimental data available. The additional studies are discussed in Sec.~\ref{sec:rover_tests}. The code used in Sec.~\ref{sec:soil_tests}~and~\ref{sec:rover_tests} is available in a GitHub public repository \cite{DEMERepoGRCBranch} for unrestricted use, distribution, and reproducibility studies. The only caveat is that the VIPER rover model shared in the public repository is only an approximation of the actual rover (in regards to its components mass, moments of inertia, wheel geometry, etc.) The shared model was put together using data available in the public domain; the actual VIPER model is not available for public access.

\subsection{Sample preparation} \label{sec:sample_prep}

\begin{figure}[htp!]
	\centering
	\captionsetup{justification=centering}
	\subfloat[The clumps need to be generated in simulation with initial gaps determined by the biggest clump to avoid overlapping. This prevents a direct generation of a densely packed granular patch.\label{fig:settle_a}]
	{
		\includegraphics[width=0.45\linewidth]{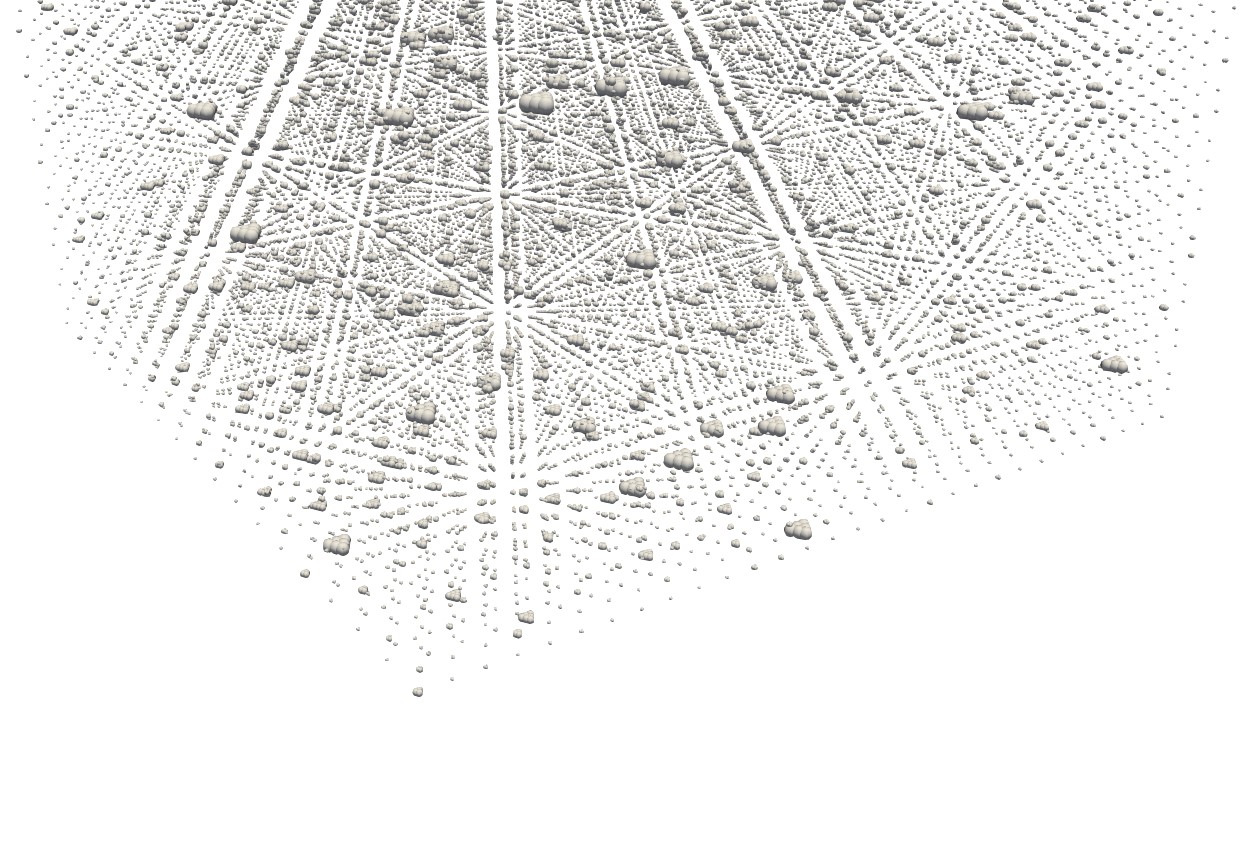}
	}
	\hspace{.1cm}
	\subfloat[The initial batch of clumps, after settling.\label{fig:settle_b}]
	{
		\includegraphics[width=0.45\linewidth]{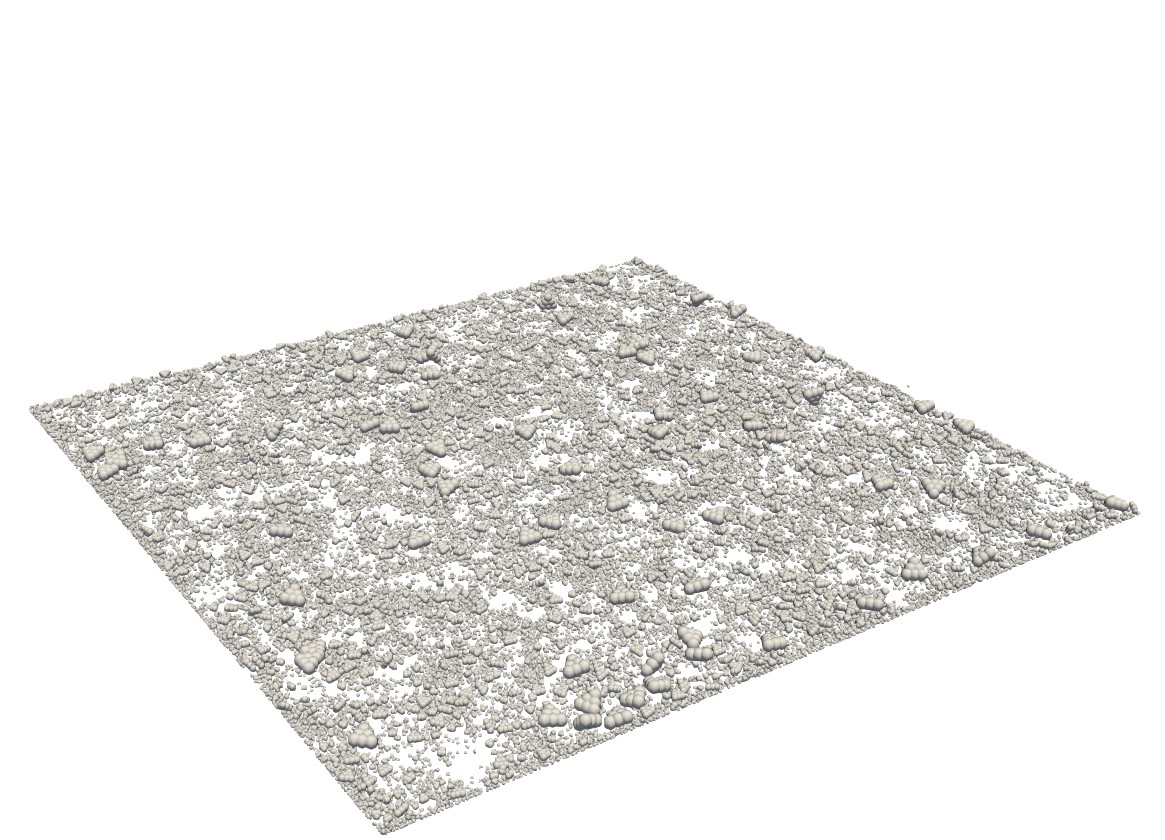}
	}
	
	\subfloat[The process is repeated and more batches of clumps are instantiated and let to be settled.\label{fig:settle_c}]
	{
		\includegraphics[width=0.45\linewidth]{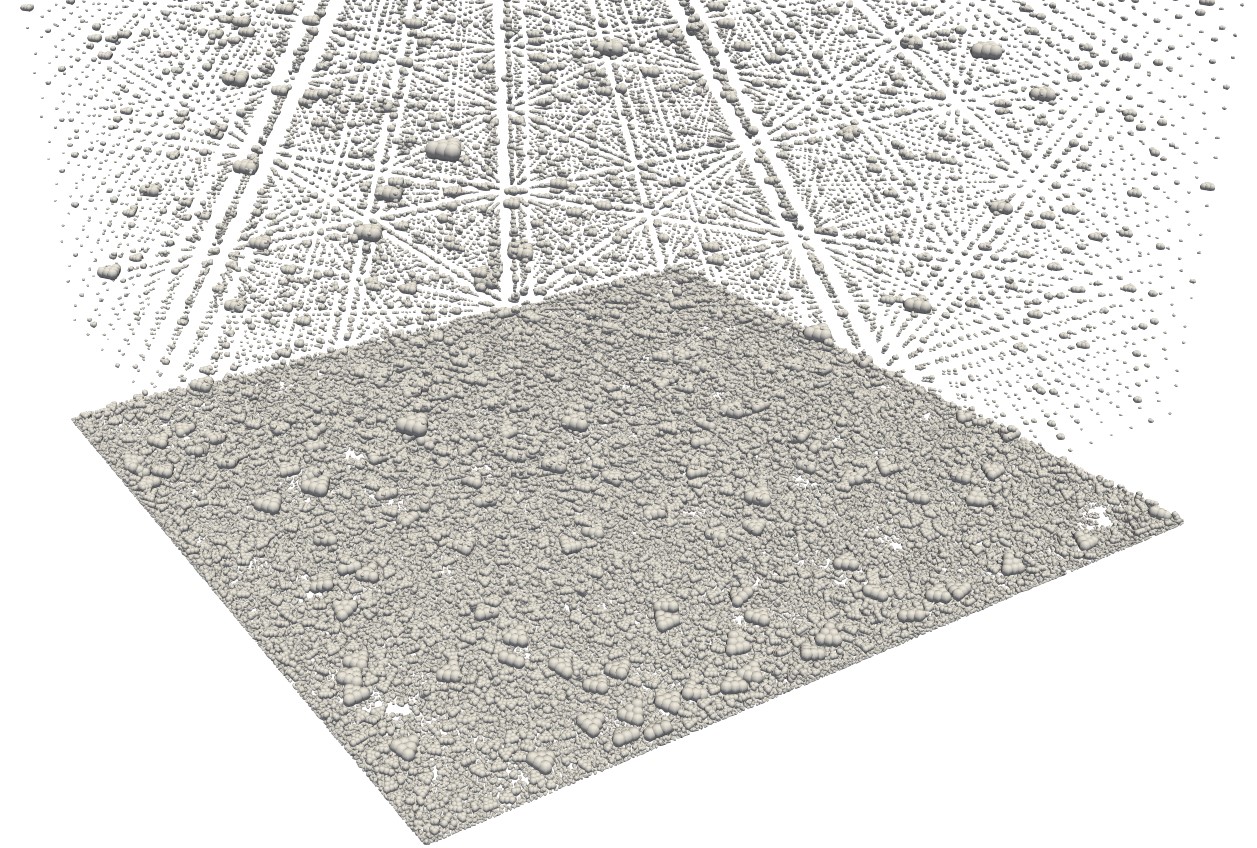}
	}
	\hspace{.1cm}
	\subfloat[The resultant granular patch after ten batches of clumps settle.\label{fig:settle_d}]
	{
		\includegraphics[width=0.45\linewidth]{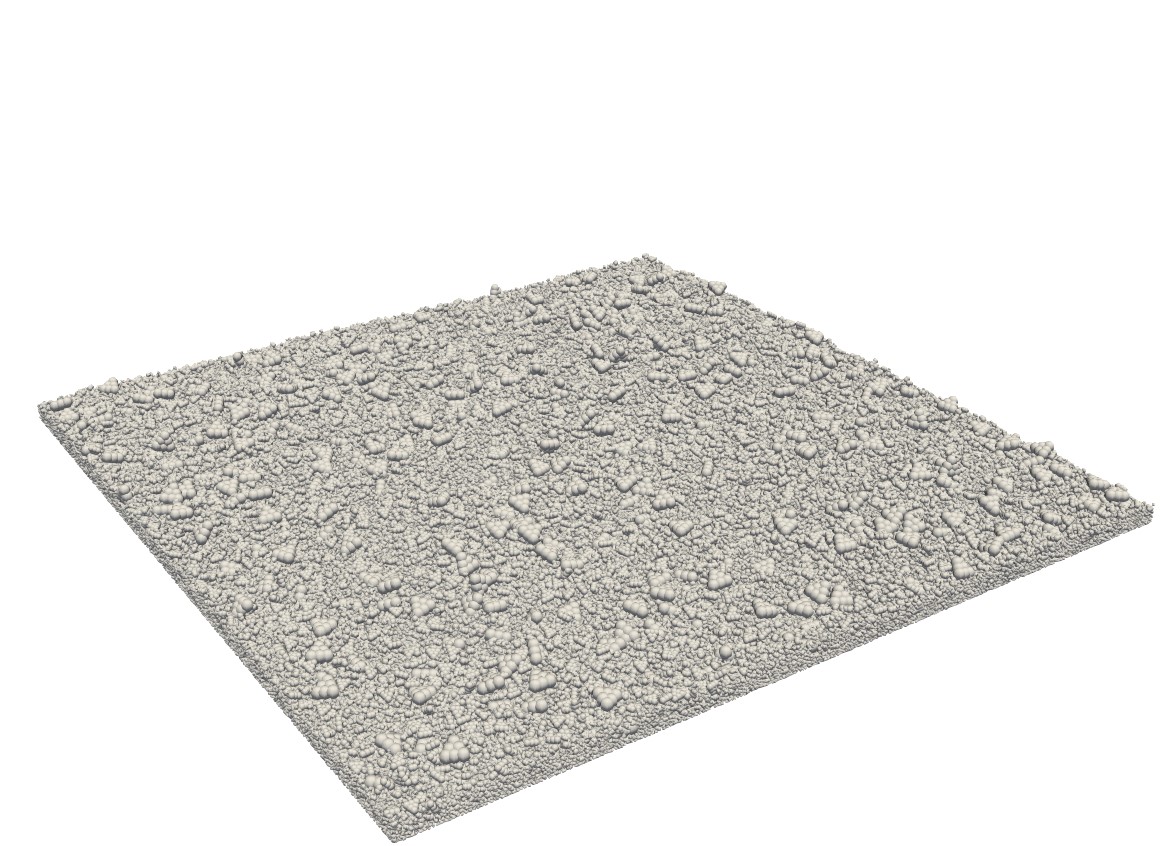}
	}
	
	\subfloat[To speed up the process, copies of the previously settled granular patch are instantiated into the simulation.\label{fig:settle_e}]
	{
		\includegraphics[width=0.45\linewidth]{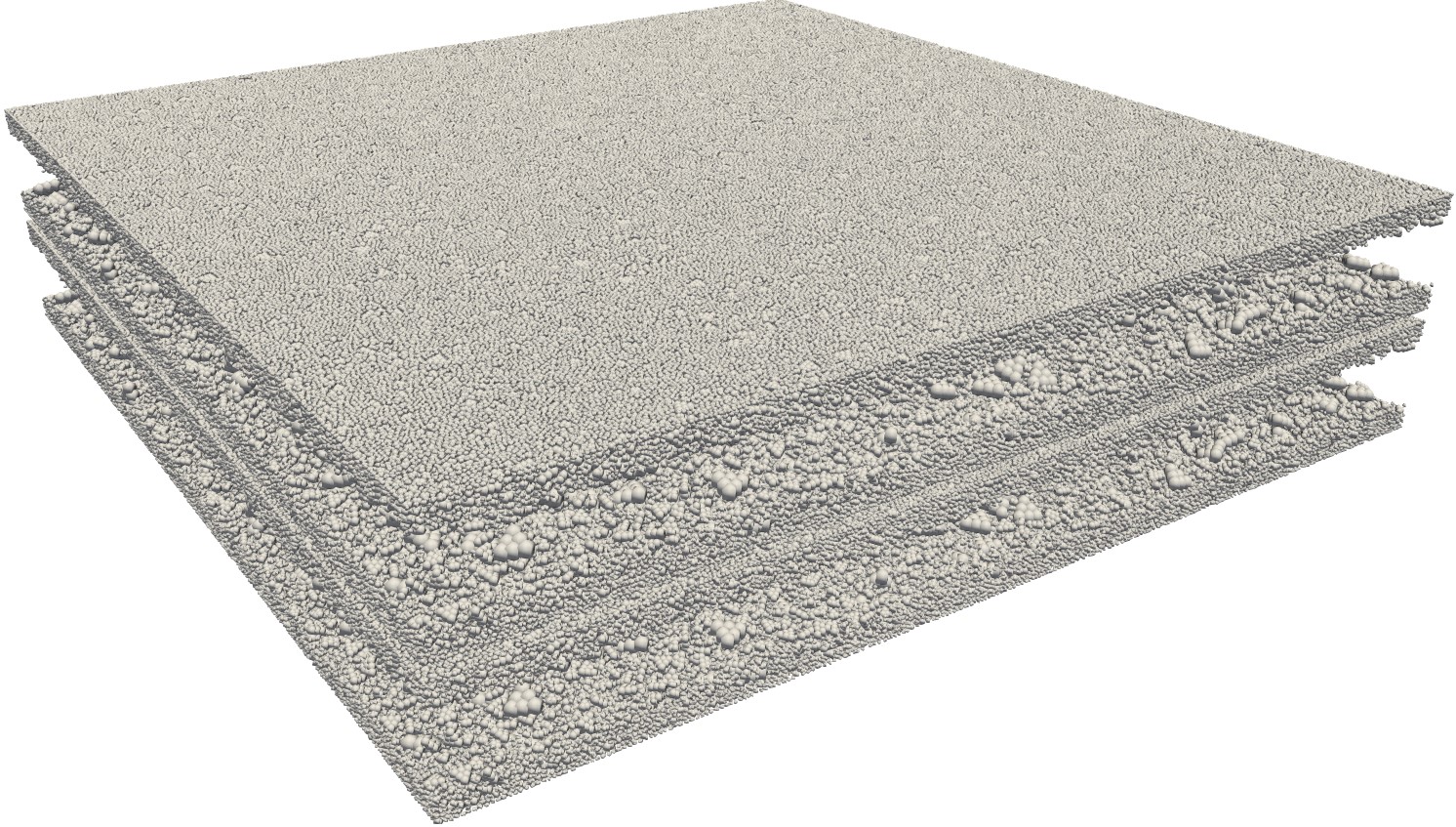}
	}
	\hspace{.1cm}
	\subfloat[The final settled and compressed base DS patch.\label{fig:settle_f}]
	{
		\includegraphics[width=0.45\linewidth]{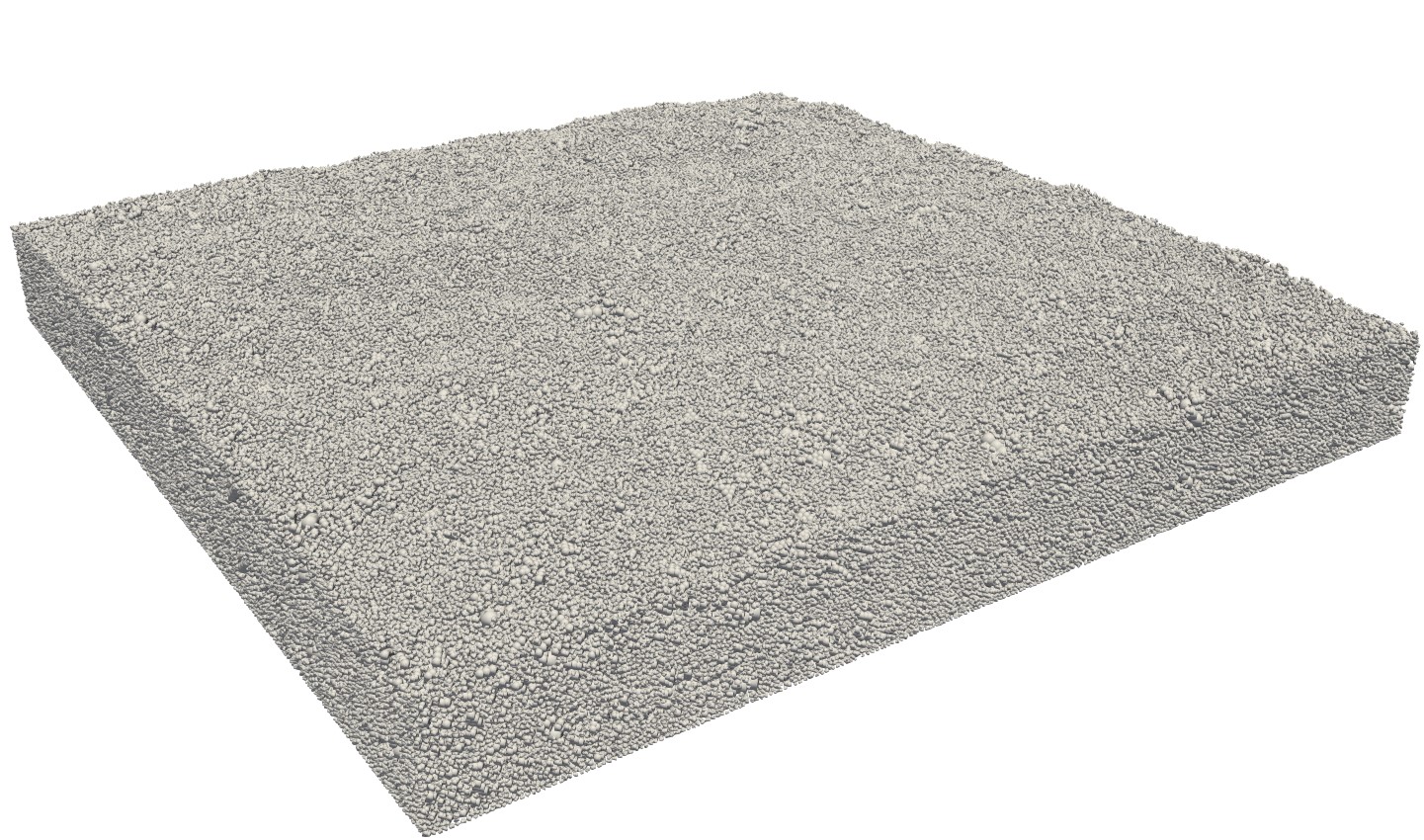}
	}
	\caption{The rendering of DS preparation process.} 
\end{figure}

This subsection describes the preparation strategy employed to produce a bed of granular material. For monodisperse spherical DEM elements, one can create the initial elements using a sampling algorithm, such as hexagonal close-packed, with a separation factor slightly larger than the diameter of each sphere. For the DS, due to the irregular grain shape and large variation in grain size, the separation factor needs to be larger than the size of the largest element to prevent overlap at the start of the simulation. This leads to a low space occupancy during the initial sampling process, as shown in Fig.~\ref{fig:settle_a}. Therefore, several batches of clumps need to be generated and settled to form a terrain sample of a certain depth, as depicted in Fig.~\ref{fig:settle_b}~and~\ref{fig:settle_c}. The element size distribution shown in Fig.~\ref{fig:GRCDS1} is retained during the spawning of each DS batch. After ten batches of clumps settle, we obtain the resultant sample illustrated in Fig.~\ref{fig:settle_d}. 

With this newly generated thin layer of clumps, we use a ``copy-paste'' process to accelerate the sample preparation. We duplicate several copies of the existing clumps, then add them back to the simulation, as shown in Fig.~\ref{fig:settle_e}. After settling, we create a virtual compressor object in the simulator and compress the sample to make the surface relatively even, see Fig.~\ref{fig:settle_f}. This terrain sample now is one meter by one meter in width and roughly 15 centimeters in height. We refer to it as a base DS patch in this paper. 
On two NVIDIA A100 GPUs, finishing the entire sample preparation process at time step size \SI{e-6}{s} takes approximately \SI{12}{} hours. There are \SI{4571136}{} clumps, totaling \SI{13993536}{} spheres that are present in the base DS patch.

The creation of this base patch is for the ease of scaling up the simulation. When a larger test bed is needed, several such patches can be instantiated, moved, and then settled in the simulator to create the test bed. Conversely, a subset of this patch can be extracted to create a smaller test bed. All numerical experiments in this paper use test beds derived from the base patch, saving time that would otherwise be spent on settling the DEM terrain.

\subsection{Repose angle validation}

\begin{figure}[htp!]
	\centering
	\captionsetup{justification=centering}
	\subfloat[The initial sample is prepared using the pre-settled patch.\label{fig:repose_a}]
	{
		\includegraphics[width=0.45\linewidth]{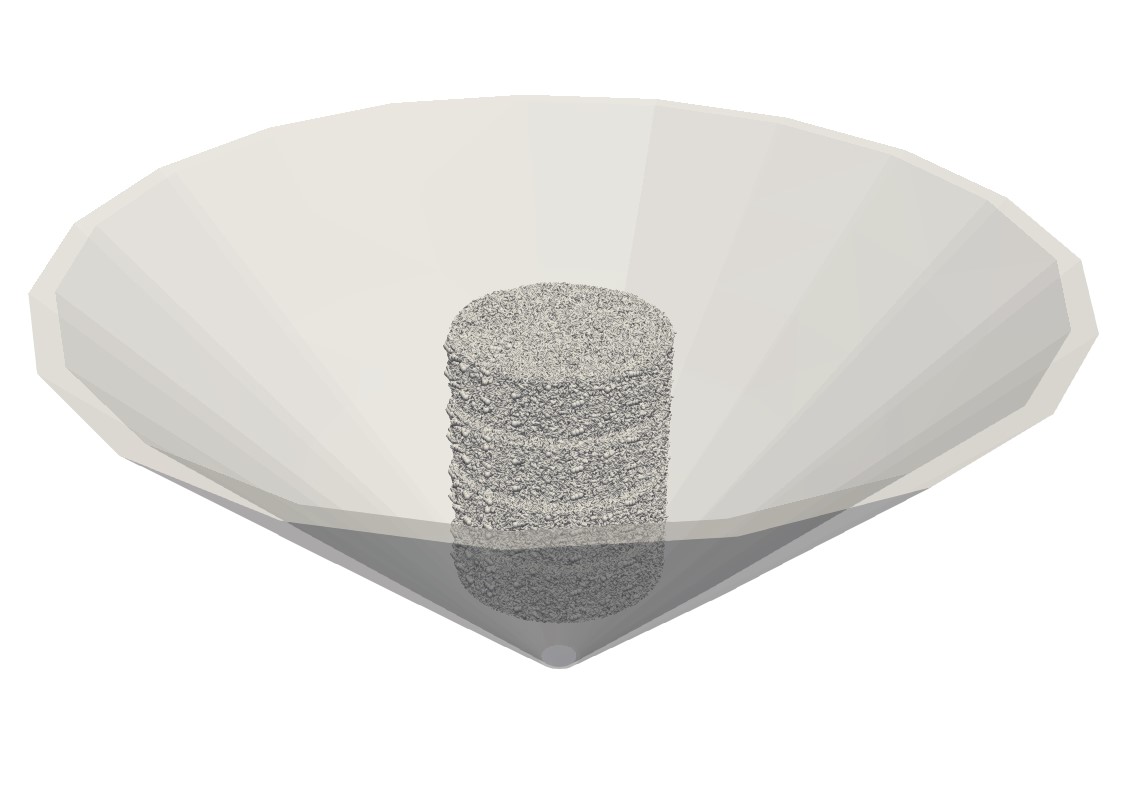}
	}
	\hspace{.1cm}
	\subfloat[A perspective view of the simulation.\label{fig:repose_b}]
	{
		\includegraphics[width=0.45\linewidth]{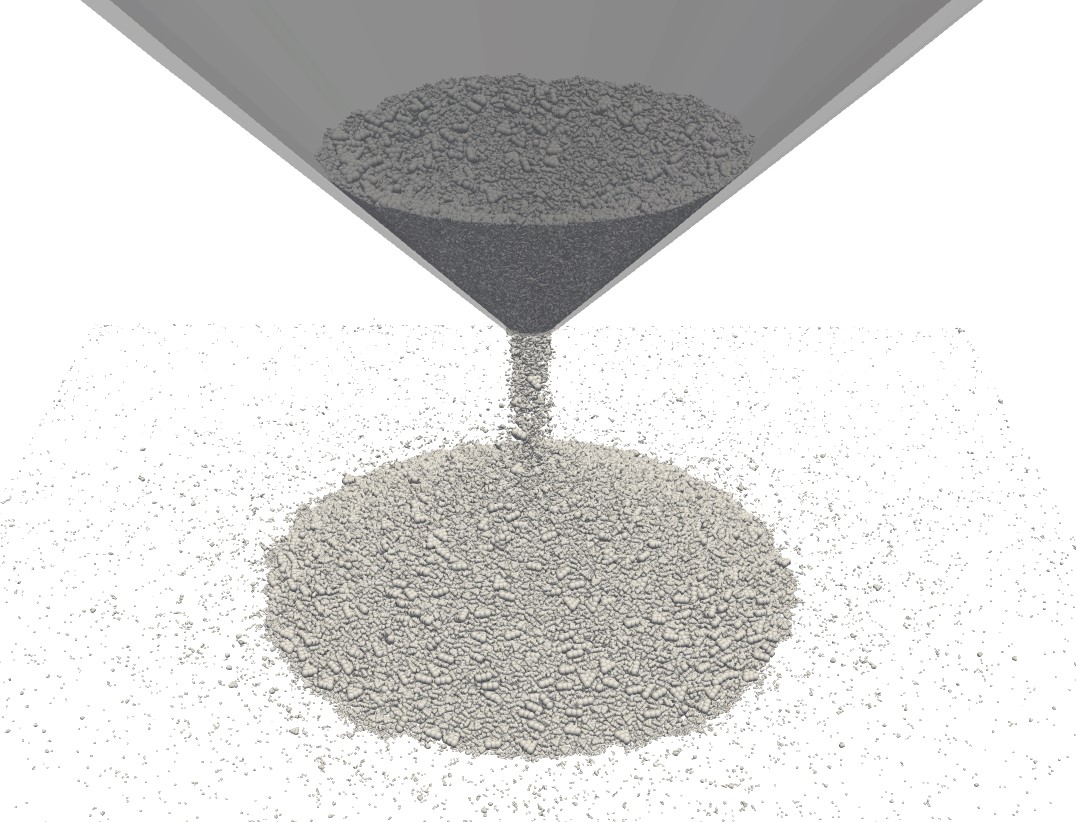}
	}
	\caption{The rendering of DS angle of repose test.} 
\end{figure}
\begin{figure}[htp!]
	\centering
	\captionsetup{justification=centering}
	\includegraphics[width=.9\linewidth]{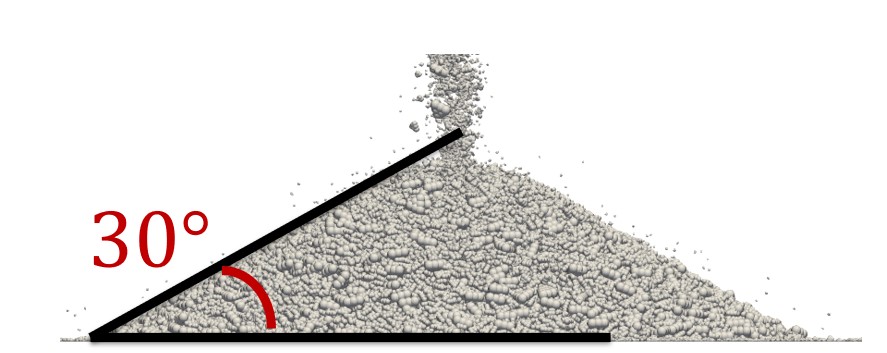}
	\caption{The settled DS forms a 30-degree pile.} \label{fig:repose_c}
\end{figure}

The angle of repose is expected to be equal to the residual internal friction angle~\cite{BEAKAWIALHASHEMI2018397}. The latter serves as a proxy for a material's shear strength that dictates a vehicle's climbing ability in off-road conditions, a topic discussed in Sec.~\ref{sec:rover_tests}. For the repose angle test, the initial sample is prepared by taking a cylindrical portion out of a DS patch, making three extra copies, and then translating everything into a funnel defined via a mesh, see Fig.~\ref{fig:repose_a}. The material flows through the funnel under gravity. The pile formed underneath can be seen in Fig.~\ref{fig:repose_b}. Figure~\ref{fig:repose_c} displays a 30 degrees angle of repose, in line with results reported in~\cite{ORAVEC2010361}. In total, \SI{731060}{} clumps participate in the simulation. 
On two NVIDIA A100 GPUs, running this \SI{30}{s} simulation at time step size \SI{e-6}{s} takes approximately \SI{19}{} hours. There are \SI{2234544}{} spheres that combine to make up the clumps in the material used for this test.

\subsection{Cone penetration test}
The cone penetration test is useful for gauging the soil's stiffness and density, as well as the stability of slopes and the potential for landslides \cite{STEINER2015162}.  Herein, we test in simulation the penetration experiments conducted with GRC-1 simulant reported in \cite{ORAVEC2010361}.
Same as in \cite{ORAVEC2010361}, the soil bin was selected to be \SI{58.4}{cm} in diameter and \SI{24}{cm} in depth, which is sufficient to minimize boundary effects \cite{Oravec2008SoilDev}.

\begin{figure}[H]
	\centering
	\captionsetup{justification=centering}
	\subfloat[The cone mesh.\label{fig:CPTa}]
	{
		\includegraphics[width=0.45\linewidth]{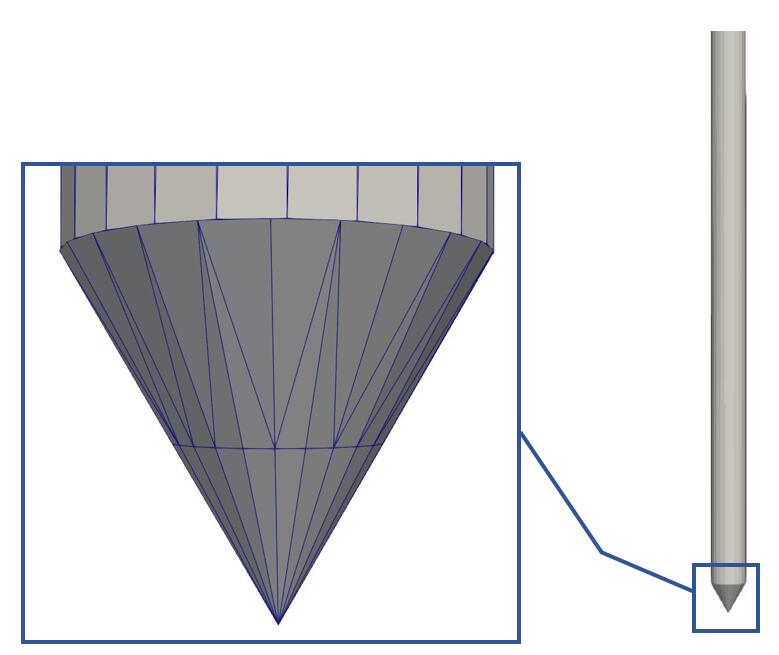}
	}
	\hspace{.1cm}
	\subfloat[A rendering of cone penetration test when the cone hits the sample.\label{fig:CPTb}]
	{
		\includegraphics[width=0.45\linewidth]{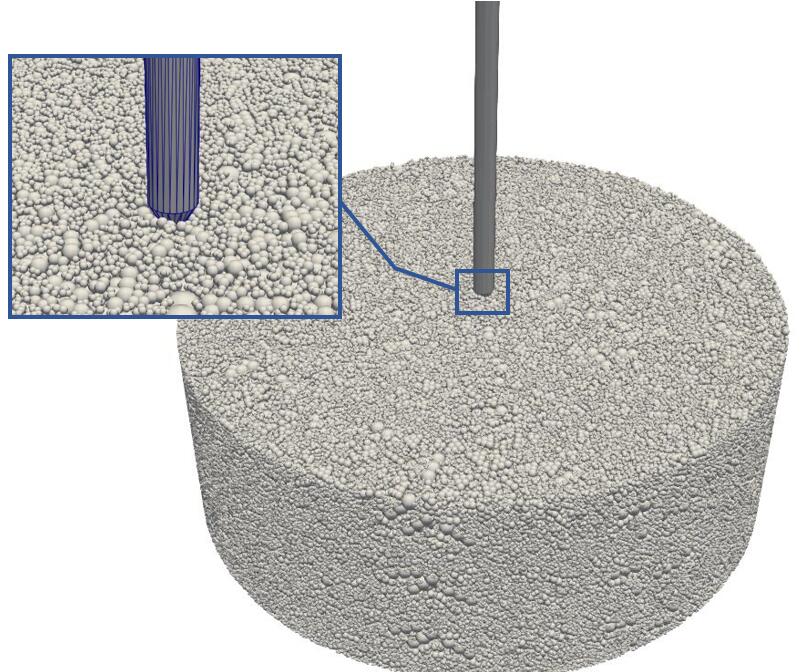}
	}
	\caption{Rendering of the cone penetration test using the DS.} \label{fig:CPT}
\end{figure}

\begin{figure}[htp!]
	\centering
	\captionsetup{justification=centering}
	\includegraphics[width=.9\linewidth]{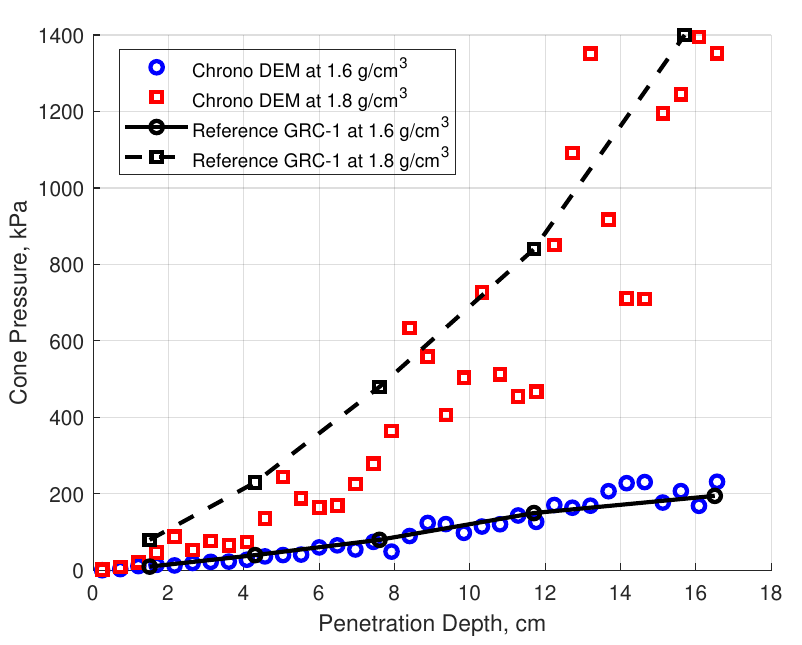}
	\caption{The DEM cone penetration test results with different bulk densities of the DS. They are plotted against the reference GRC-1 cone penetration experimental data. The experimental data are from GRC-1 simulant research paper~\cite{ORAVEC2010361}.} \label{fig:CPT_res}
\end{figure}

Similar to the repose test,  the initial sample is prepared by taking a cylindrical portion out of a DS patch, making two extra copies, and then translating everything into the cylindrical container and letting the material settle.
Subsequently, a cone penetrates the sample with a constant velocity of \SI{3}{cm/s}, see Fig.~\ref{fig:CPTb}. The cone, whose geometry is specified by a mesh, is shown in Fig.~\ref{fig:CPTa}. The cone has a base area of \SI{323}{mm^2} and an opening angle of $60^\circ$. The contact force on the cone is measured and plotted against the penetration depth in Fig.~\ref{fig:CPT_res}. The different bulk density densities of DS in  Fig.~\ref{fig:CPT_res} are reached by compressing the material bed before the simulation starts.

As shown in Fig.~\ref{fig:CPT_res}, the DS reproduces a depth--pressure relationship in the cone penetration tests similar to the one noted in real-world experiments \cite{ORAVEC2010361}.
The simulator also reproduces the drastic increase of the pressure on the cone when the simulant is compressed to a higher bulk density, from \SI{1.6}{g/mm^3} to \SI{1.8}{g/mm^3}. In all the subsequent numerical tests presented in the paper, our DS operates at a bulk density around  \SI{1.6}{g/mm^3}, which is close to what a naturally settled DS sample has. 

In total, \SI{773097}{} clumps, which employ \SI{2362698}{} spheres, participate in the simulation. On two NVIDIA A100 GPUs, running this  \SI{8}{s} simulation at time step size \SI{e-6}{s} takes approximately \SI{4}{} hours.

\subsection{Single-wheel drawbar pull test}

In \cite{Senatore2014ModelingAV}, the authors used a test rig ARTEMIS to derive the force--slip relationship for a single Curiosity wheel operating on Mars soil simulant. We used the experimental setup described in their paper as a reference and demonstrated that our DS captures the terrain strength similar to existing simulants. Note that there is a slight discrepancy between the actual granular material used in \cite{Senatore2014ModelingAV} and the DS used in the simulations. This aspect will be revisited shortly.

Reproducing drawbar pull tests in simulation involves firstly an abstraction of the test equipment. The terrain leveling and compacting process described in~\cite{Creager2017DrawbarP} has already been incorporated into the terrain preparation procedure discussed in Sec.~\ref{sec:sample_prep}. Additionally, enforcing a slip value does not require a mechanical system or attaching a hitch to the test vehicle as in the physical test; instead, specific linear and angular velocities are applied directly to the wheel.

\begin{figure}[htp!]
	\centering
	\captionsetup{justification=centering}
	\subfloat[Open source mesh of the Curiosity wheel, publicly available in the Chrono repository~\cite{projectChronoWebSite}.\label{fig:DP_a}]
	{
		\includegraphics[width=0.35\linewidth]{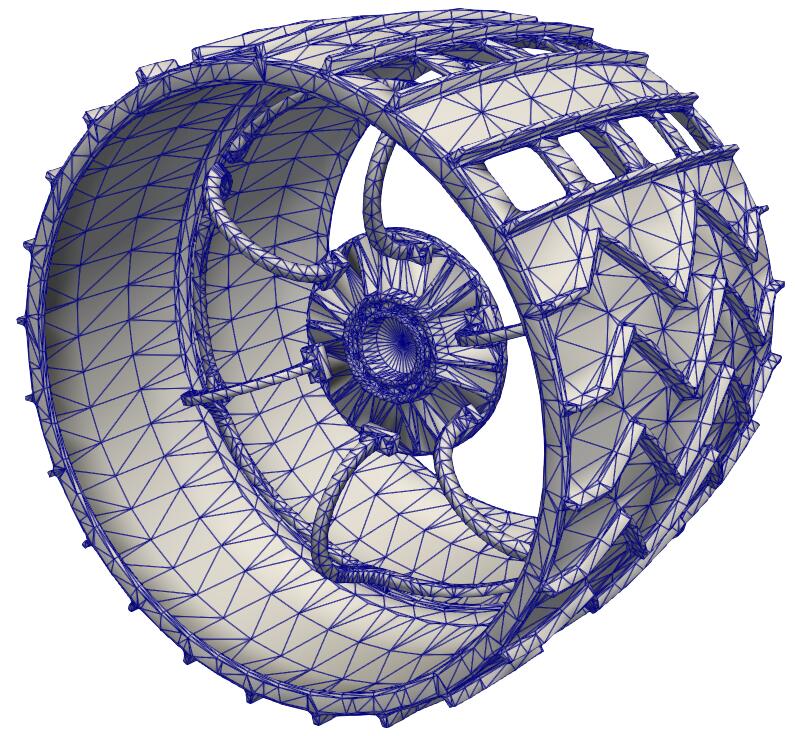}
	}
	\hspace{.03cm}
	\subfloat[A perspective view of the wheel operating on the soil bed.\label{fig:DP_b}]
	{
		\includegraphics[width=0.6\linewidth]{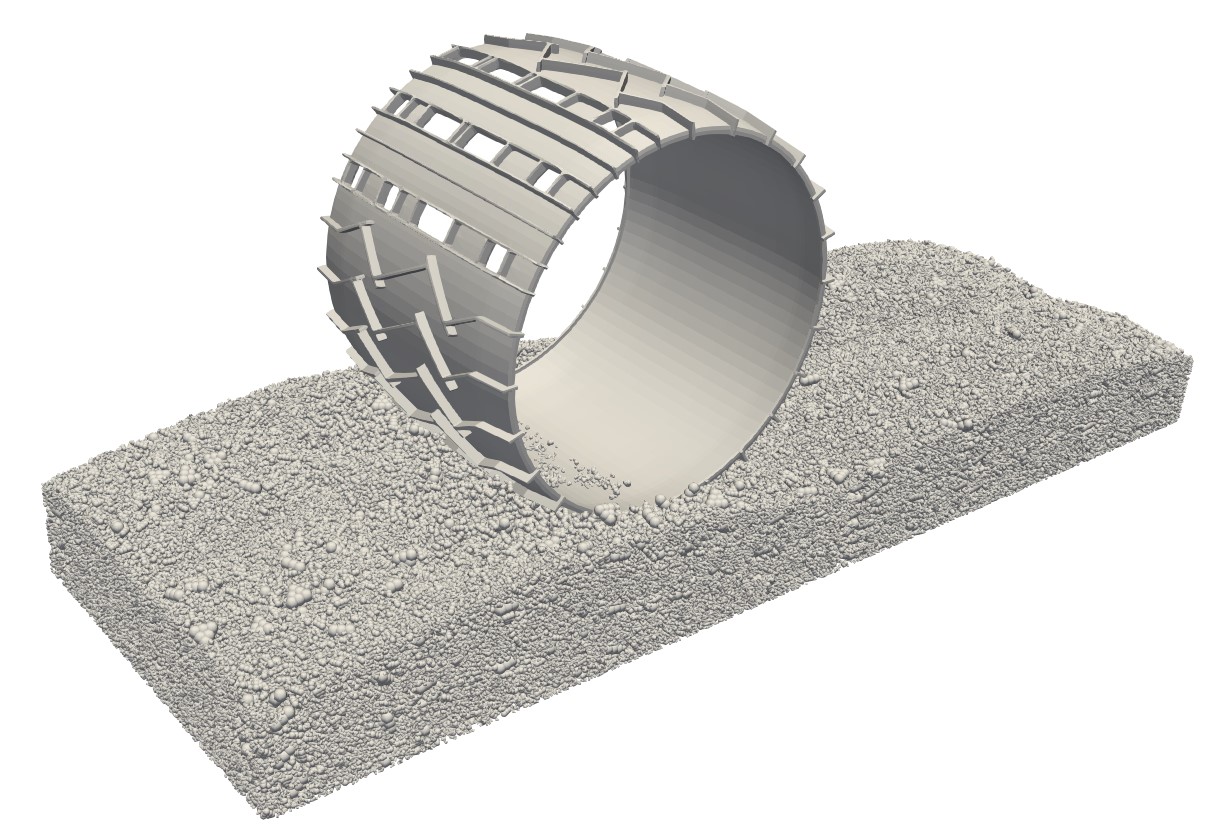}
	}
	\caption{The mesh and the simulation scene of the drawbar pull test.}
\end{figure}

In this test we use Earth's gravitational acceleration is $g=\SI{9.81}{m/s^2}$. The Curiosity wheel is represented by the mesh shown in Fig.~\ref{fig:DP_a}. The grouser geometry is preserved exactly; the wheel has a radius of $r = \SI{0.25}{m}$ and a weight of \SI{750}{N}. The mesh file can be obtained from the Chrono repository~\cite{projectChronoWebSite}, and readers are referred to \cite{Senatore2014ModelingAV} for comparison against the real Curiosity wheel. In this simulation, the wheel hub is removed to reduce the mesh size and save simulation time.

An overview of the simulation environment is illustrated in Fig.~\ref{fig:DP_b}. The granular terrain is prepared by making copies of the DS patch and concatenating these copies together so that a test bed is formed. A fixed angular velocity of $\omega = \SI[parse-numbers = false]{(\pi/12)}{rad/s}$ is imposed on the wheel. The linear velocity $v$ in the forward direction is enforced as $v = (1-s)\omega  r$, where $s$ is the desired slip ratio. 

\begin{figure}[htp!]
	\centering
	\captionsetup{justification=centering}
	\includegraphics[width=.9\linewidth]{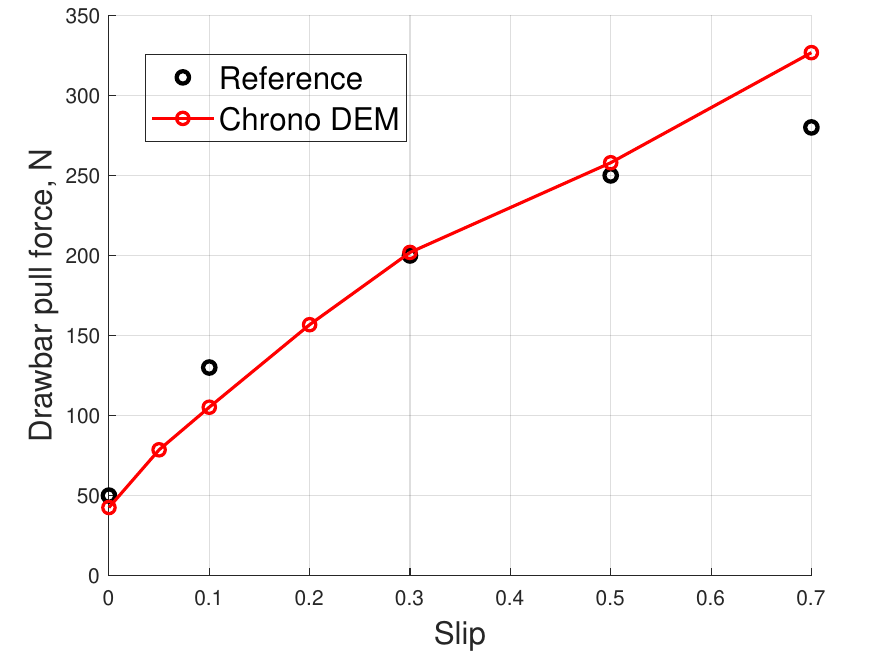}
	\caption{Curiosity wheel drawbar pull DEM simulation result with a vertical load of \SI{750}{N}, compared against the reference ARTEMIS experimental data~\cite{Senatore2014ModelingAV}.} \label{fig:DP_res}
\end{figure}

The simulation results are presented in Fig.~\ref{fig:DP_res}. 
The reference experimental data from \cite{Senatore2014ModelingAV} is shown with black markers in the same plot. Note that the simulations in \cite{Senatore2014ModelingAV} use a custom-made blend of two types of sand that is different from GRC-1, as we do not have access to the drawbar pull data of GRC-1. Therefore, this comparison is qualitative only. Nonetheless, our drawbar pull--slip curves exhibit notable similarities between our simulation and the reference experiment.

In total, \SI{960906}{} clumps employing \SI{2942346}{} spheres participate in the simulation. On two NVIDIA A100 GPUs, running this  \SI{8}{s} simulation at time step size \SI{e-6}{s} takes approximately \SI{5}{} hours.

\section{Further DEM studies related to rover mobility} \label{sec:rover_tests}
This section focuses on single-wheel and full-rover incline tests. The limited amount of experimental data used for validation was produced by the Simulated Lunar Operations (SLOPE) laboratory at NASA's Glenn Research Center using GRC-1 and GRC-3 simulants. The wheel geometry used in the tests of this section is from NASA's Moon Gravitation Representative Unit 3 (MGRU3).

\subsection{Single-wheel slip test on incline} \label{sec:sw}

\begin{figure}
	\centering
	\captionsetup{justification=centering}
	\subfloat[A $\alpha$-degree ``incline'' in DEM simulations is created by tilting the direction of the gravity.\label{fig:wheel_a}]
	{
		\includegraphics[width=0.45\linewidth]{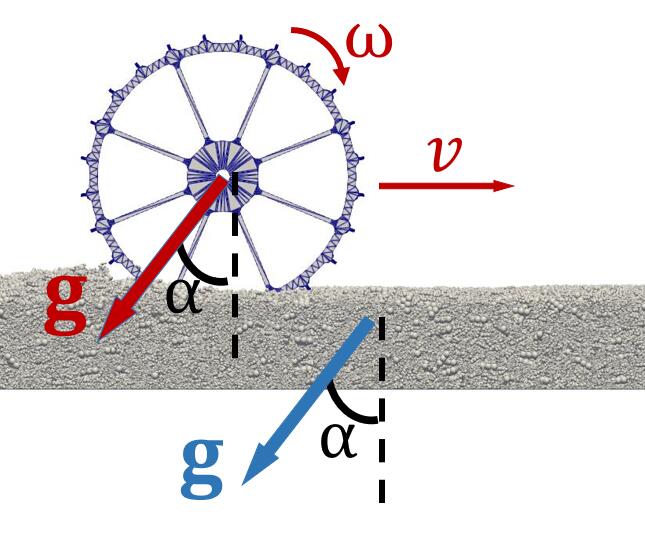}
	}
	\hspace{.1cm}
	\subfloat[A rendering of single-wheel test scene.\label{fig:wheel_b}]
	{
		\includegraphics[width=0.45\linewidth]{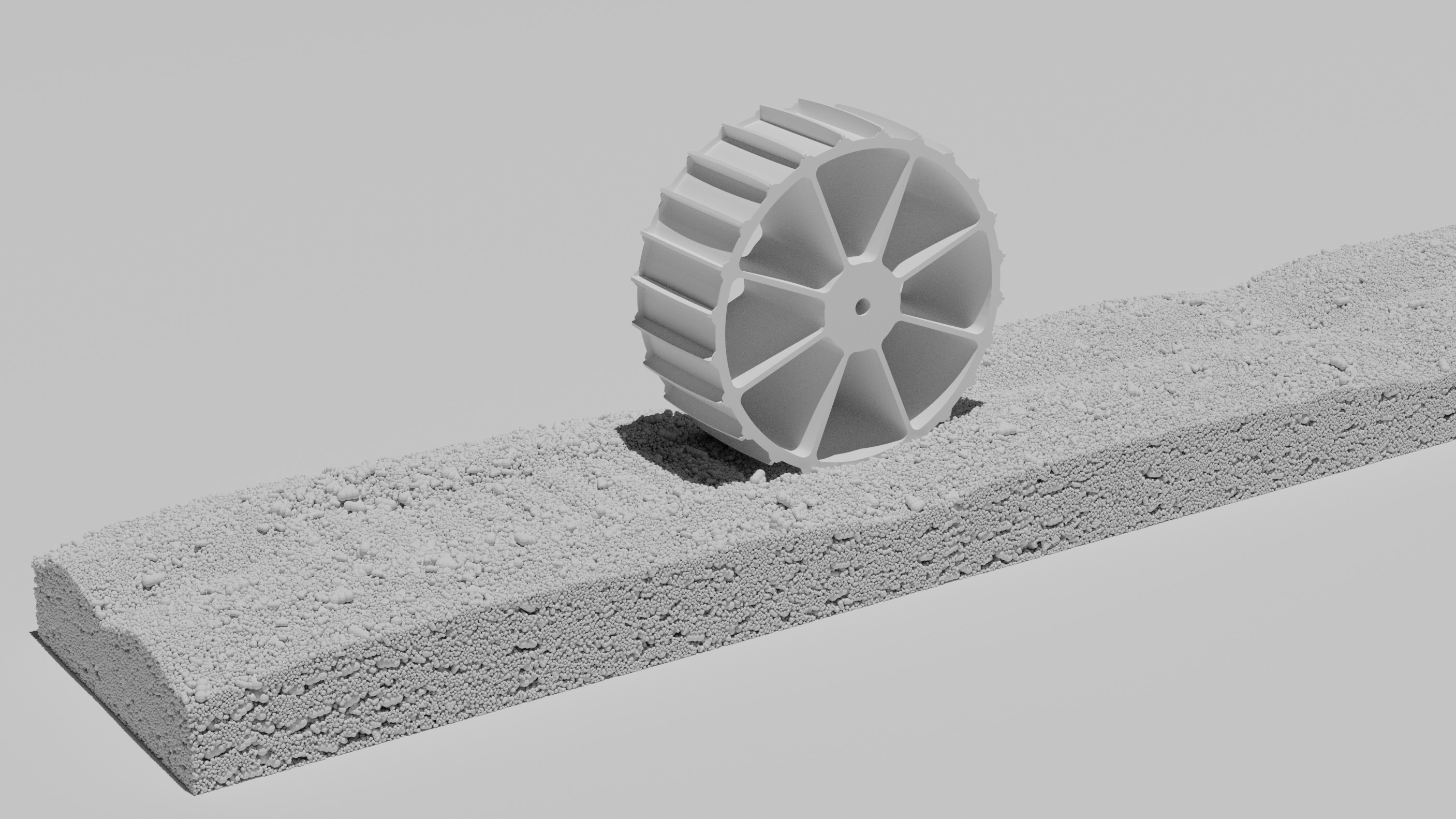}
	}
	\caption{The simulation setup for DEM single-wheel tests using DS.}
\end{figure}

The inclines used in the simulation are created by altering the direction of gravity, see Fig.~\ref{fig:wheel_a}. This approach to representing slopes requires minimum model adjustments when switching from one test scenario to another.
Figure~\ref{fig:wheel_b} is a screenshot of this test. 
The wheel in the test is subject to a fixed angular velocity $\omega$ and is free to move linearly. The steady-state average linear velocity over 6 seconds $v$ is measured, and subsequently used to obtain the slip ratio $s=1-v/(\omega r)$.
The wheel used has a mass of \SI{5}{kg} but may have a different ``effective'' mass in these simulations. The ``effective'' mass is introduced to account for the extra mass placed on top of each rover wheel. This modified mass is modeled as an extra gravity-aligned force applied to the center of the wheel. This force is dependent on the gravitational acceleration of the test scene.
For example, if a wheel has an effective mass of \SI{22}{kg} in a test that uses the Moon's gravity, that means an extra force of \SI[parse-numbers = false]{17\times 1.62}{N} is applied to it. 
The granular terrain is prepared by making copies of the base DS patch and concatenating these copies together, similar to the previous drawbar pull test.

\begin{figure}[htp!]
	\centering
	\captionsetup{justification=centering}
	\includegraphics[width=\linewidth]{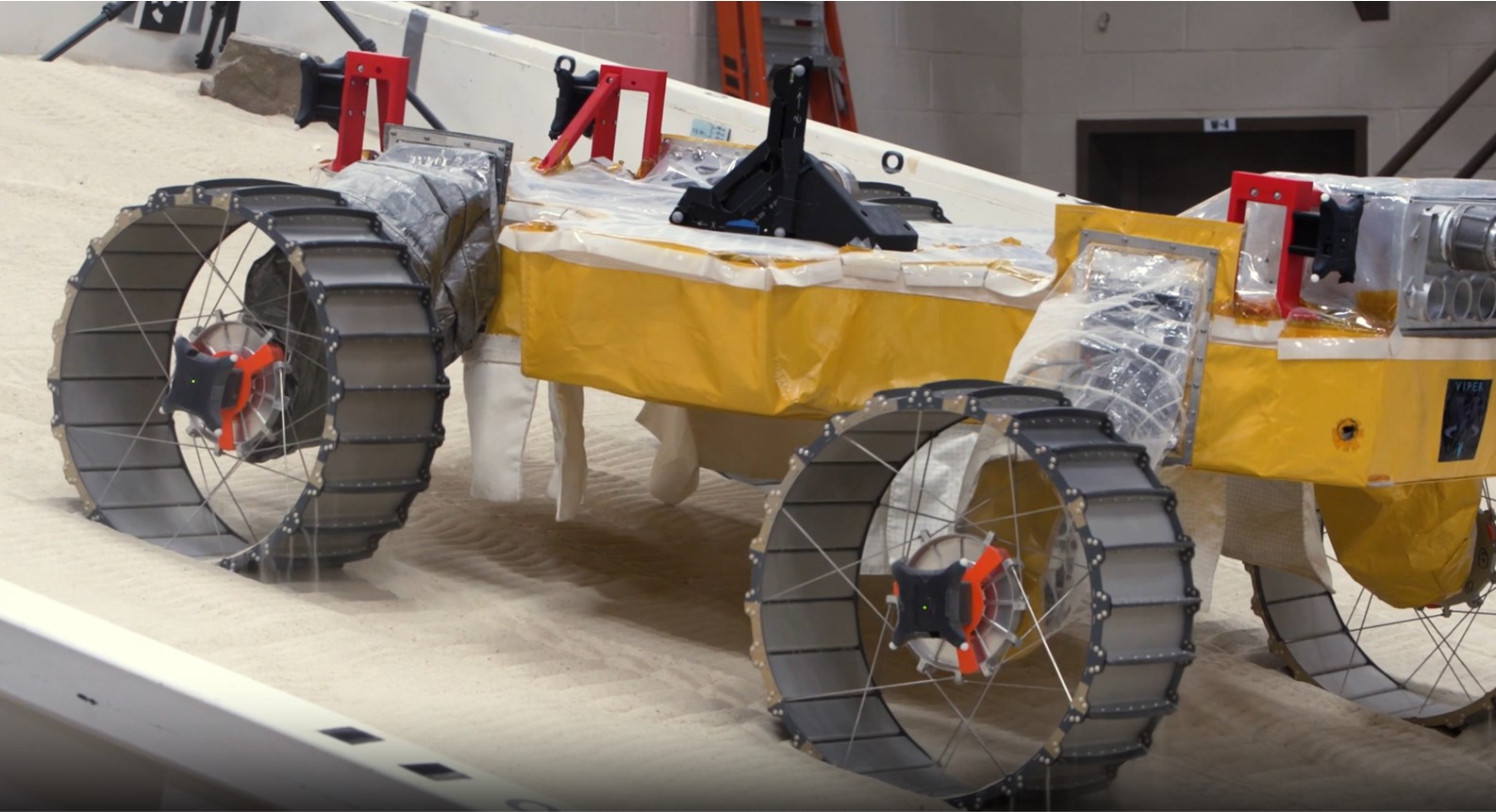}
	\vspace{5pt}
	\caption{MGRU3 climbing a ``tilt bed'' in NASA's SLOPE lab testing facility.}\label{fig:MGRU3movie}
	\vspace{-10pt}
\end{figure}

The ground-truth experimental data used for comparison is from the experiments done with MGRU3 climbing a ``tilt bed'' made with the GRC-1 simulant in the SLOPE lab. A picture of the test scene can be seen in Fig.~\ref{fig:MGRU3movie}, which is from a publicly available video of the test~\cite{movieMGRU3Tilt}.  In this experiment, the wheels of MGRU3 were operating at $\omega'=\SI{0.8}{rad/s}$ under Earth's gravity $g=\SI{9.81}{m/s^2}$. Note that the thickness of the bed of granular material has an influence on the experimental results, and the sufficient thickness to eliminate the boundary effect could depend on the vertical load. However, this secondary aspect, which might produce slightly different data in both experiments and simulations, falls outside the scope of this contribution.

\begin{figure}[htp!]
	\centering
	\captionsetup{justification=centering}
	\subfloat[Slip-on-incline test result with \SI{22}{kg} wheel. The experimental data used for comparison (black line) are from Glenn Research Center's MGRU3 experiments with the GRC-1 simulant.\label{fig:wheel_res_a}]
	{
		\includegraphics[width=.9\linewidth]{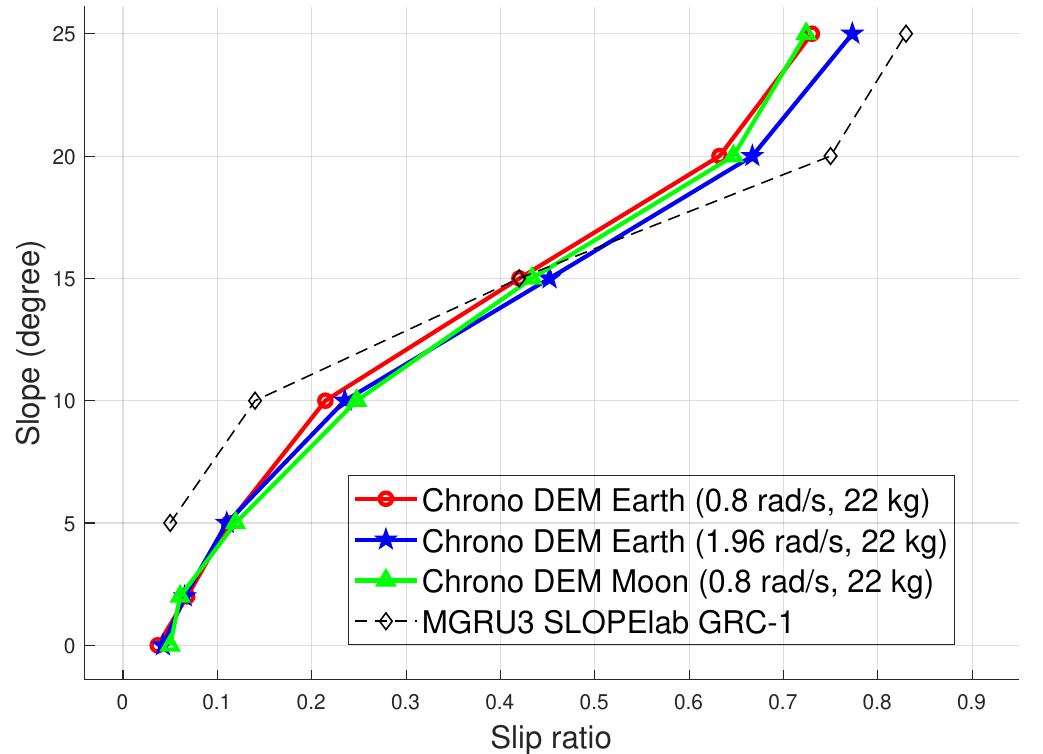}
	}
	\vspace{.1cm}
	
	\subfloat[Slip-on-incline test result with \SI{111}{kg} wheel. The experimental data used for comparison (black line) are from ProtoInnovations' single-wheel experiments with the  GRC-3 simulant.\label{fig:wheel_res_b}]
	{
		\includegraphics[width=.9\linewidth]{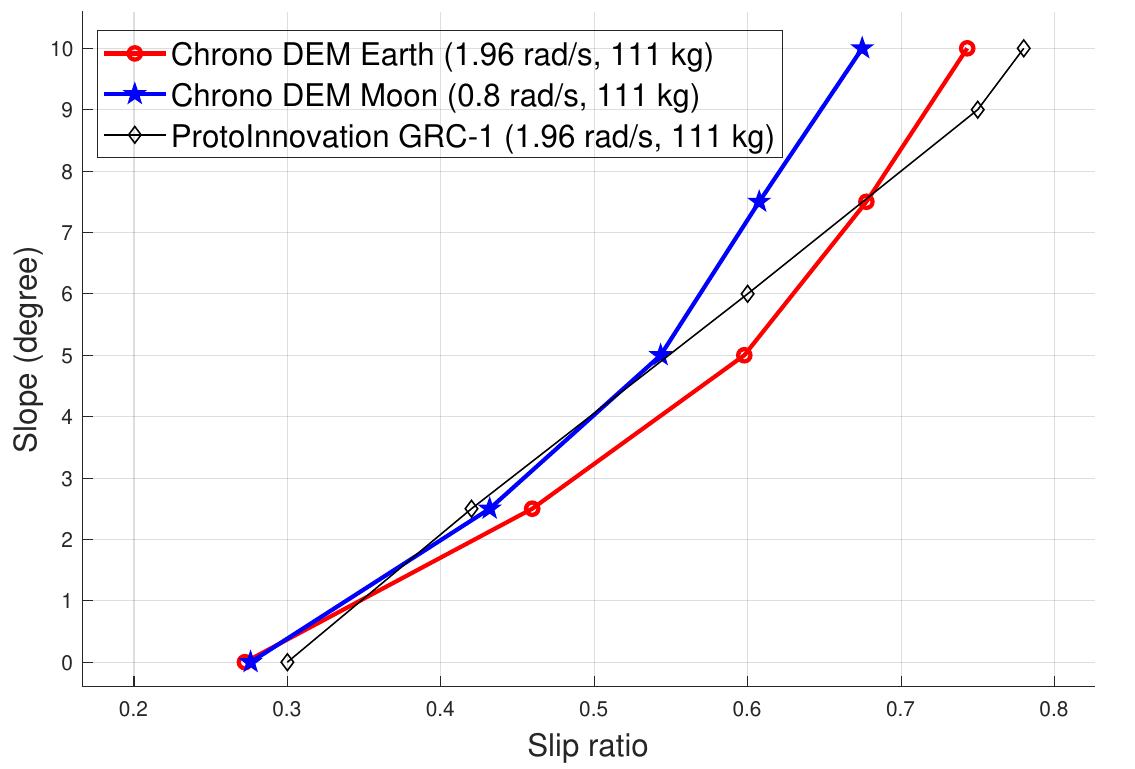}
	}
	\caption{The single-wheel slope vs. slip DEM test results with a variety of wheel effective mass, angular velocity, and gravitational pull.} 
\end{figure}

\begin{figure}[htp!]
	\centering
	\captionsetup{justification=centering}
	\includegraphics[width=.9\linewidth]{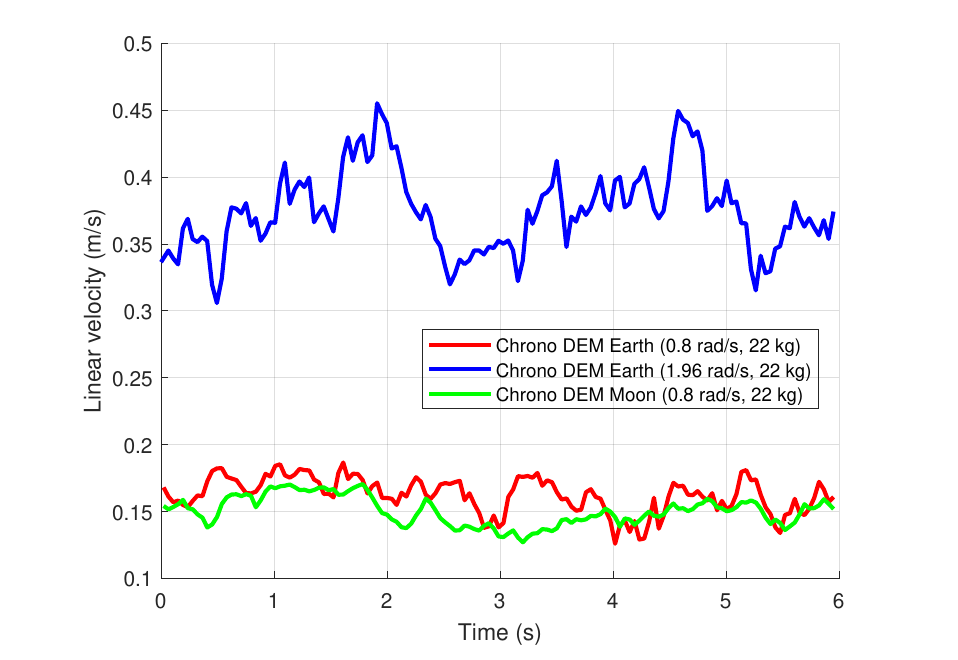}
	\caption{The linear velocity of the wheel along the incline direction plotted against the 6-second time span that we used to derive the slip ratio, at a $10^\circ$ incline.} \label{fig:velocity_series}
\end{figure}

Fig.~\ref{fig:wheel_res_a} gives the slope--slip relationship for a wheel with an effective mass of \SI{22}{kg}. 
Our first DEM experiment is set to match the experimental baseline test condition. The result is shown with the red curve. Two additional single-wheel tests are subsequently conducted: the blue curve shows the result of a test with angular velocity $\omega=\SI[parse-numbers = false]{1.96}{rad/s}$ under Earth's gravity; the green curve
is associated with a test in with angular velocity $\omega'=\SI[parse-numbers = false]{0.8}{rad/s}$ under moon gravity $g'=\SI[parse-numbers = false]{1.62}{m/s^2}$.
The similarity of the results indicates that the simulator captures the scaling law for locomotion in granular media \cite{slonaker2016geometrically,PhysRevE95052901}, which states that, if $g'/\omega'^2=g/\omega^2$, the slip ratio at steady state is unchanged.
At the same time, the fact that they are also close to the red curve suggests the slip ratio is relatively insensitive to the variation of the rotational speed, at least within the range we tested. 
These three test scenarios demonstrate that it is viable to reproduce the terrain response on the Moon by using the same simulant on Earth while keeping the rover \emph{mass} constant. 
Using the 10-degree simulation as an example, we plot the linear velocity of the wheel along the incline direction against the 6-second time span in which we measure the average velocity $v$ in Fig.~\ref{fig:velocity_series}. The linear velocity shows a well-defined mean value, meaning a steady state is reached during our measurement.

Next, we use another single-wheel test shown in  Fig.~\ref{fig:wheel_res_b} to demonstrate that matching the \emph{ground pressure} under a wheel on the Moon and Earth is not an accurate approach for predicting the traction capability of a rover.
The wheel in this test has an effective mass of \SI{111}{kg}. 
If matching the ground pressure on Earth would reproduce the rover behavior on the Moon, then we would expect the blue curve in  Fig.~\ref{fig:wheel_res_b} to show a similar slip ratio at a given slope, compared to the red curve in  Fig.~\ref{fig:wheel_res_a}.
However, this is not the case. To negotiate the same slope, a lot more slip is observed in this \SI{111}{kg} test because the terrain elements also shear a lot more easily in the low-gravity condition.
Note that the scaling law holds in this test condition too, as shown by the similarity between the two curves in this plot.
The ground-truth experimental data used for comparison (black dashed line) are from  ProtoInnovations' single-wheel experiments with the GRC-3 simulant, because the experimental data for \SI{111}{kg} wheel operating on GRC-1 simulant is not available. Note that GRC-3 simulant has in general comparable properties to GRC-1, but contains slit particles, and therefore this comparison is qualitative only. Those experiments were done using  ProtoInnovations' in-house test bed, with the wheel operating at $\SI{1.96}{rad/s}$ under Earth's gravity $g=\SI{9.81}{m/s^2}$.

A number of \SI{3681560}{} DEM elements employing \SI{11259462}{} spheres are used in the simulations for which $\omega=\SI{1.96}{rad/s}$. On two NVIDIA A100s, running these  \SI{8.4}{s} simulations at time step size \SI{e-6}{s} took approximately \SI{15}{} hours. For test cases where the angular velocity of the wheel is smaller, shorter test beds are used and the time cost is proportionally lower.

The simulation results obtained in this subsection are in agreement with the general scaling relation for locomotion in granular media \cite{PhysRevE95052901} and they underline an important fact: to carry out single wheel tests on Earth that display slope vs. slip curves similar to the ones obtained on the Moon, one should keep the same mass of the wheel. The remaining single-wheel simulations carried out in this section employ the \SI{22}{kg} wheel and look into the sensitivity of the slope vs. slip curves with respect to three factors: the size of the clumps that make up the terrain, value of the friction coefficient, and clump shape. 

\subsubsection{Sensitivity of slope vs. slip curve with respect to the clump size}\label{sec:sen_size}

\begin{figure}[htp!]
\centering
\begin{minipage}{.44\textwidth}
	\centering
	\captionsetup{justification=centering}
	\includegraphics[width=\linewidth]{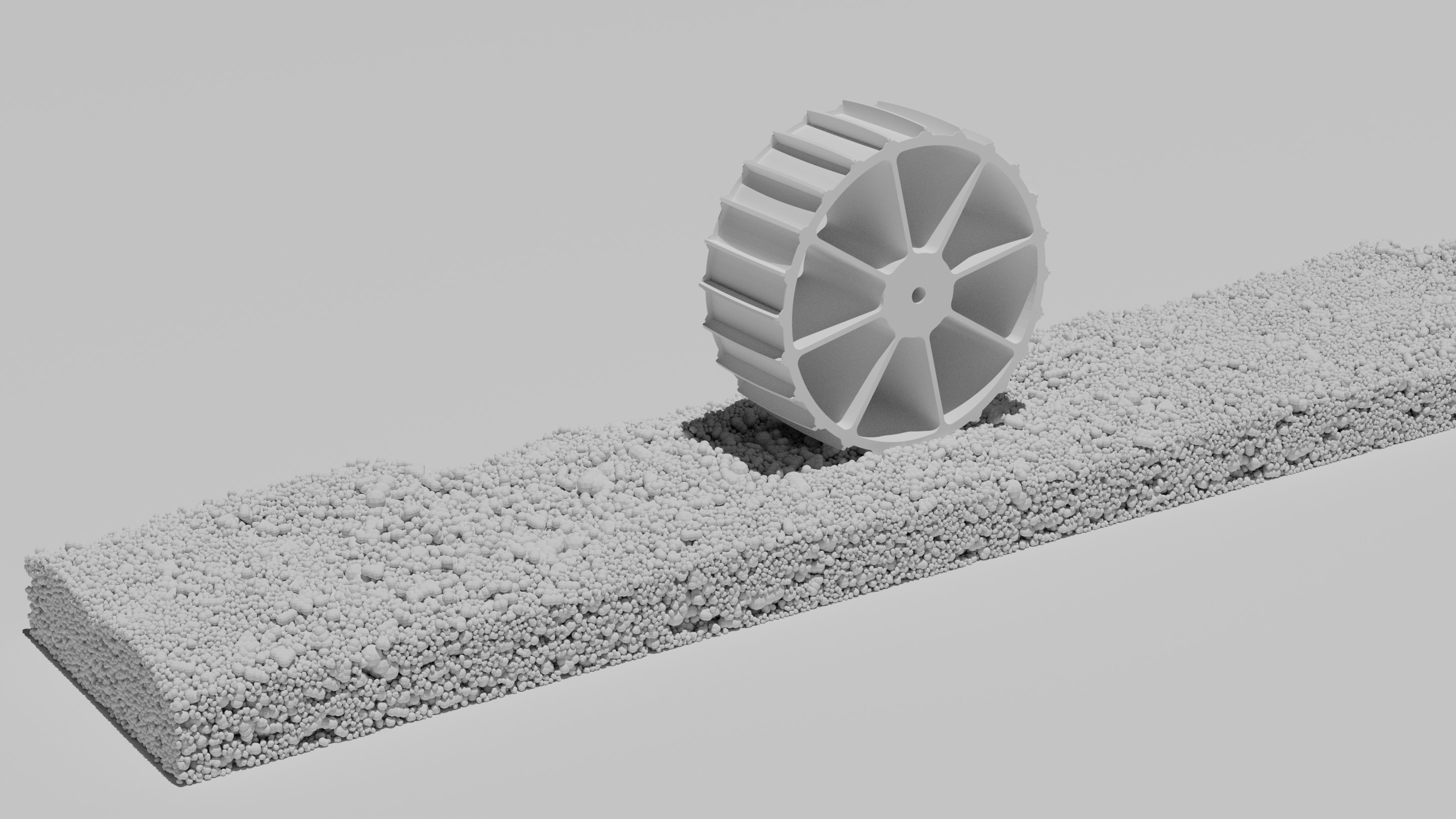}
	\caption{A rendering of the single-wheel test with enlarged terrain clumps.} \label{fig:sph_wheel_bigger}
\end{minipage}%
\hspace{.08cm}
\begin{minipage}{.52\textwidth}
	\centering
	\captionsetup{justification=centering}
	\includegraphics[width=\linewidth]{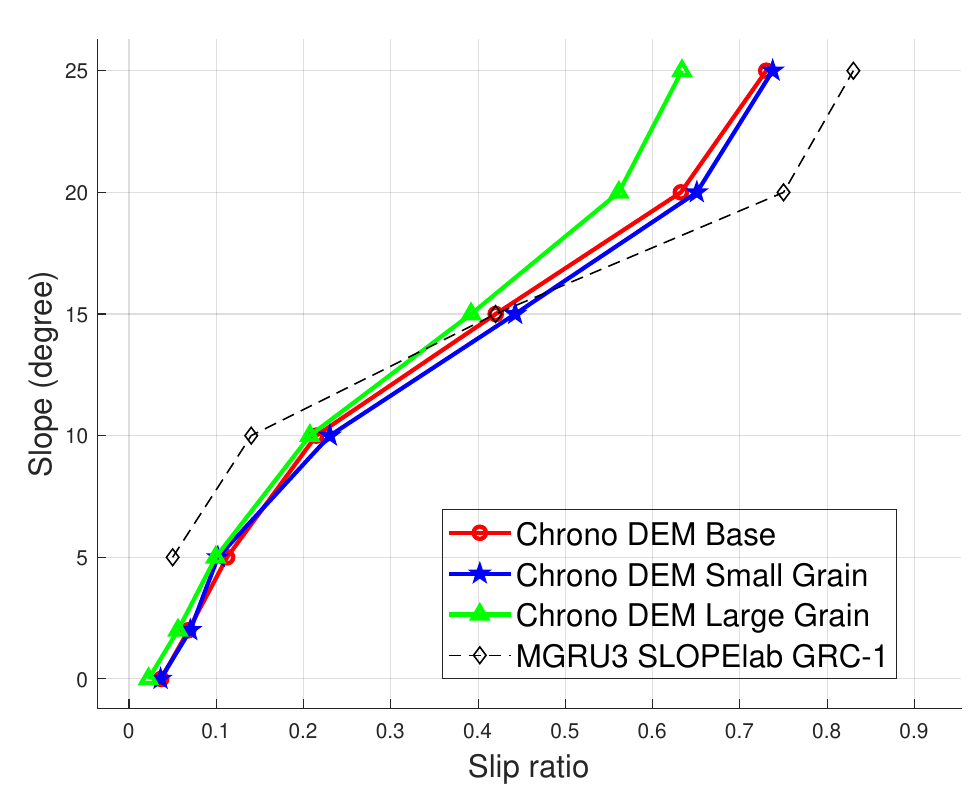}
	\caption{The single-wheel slope--slip test results with different selections of the clump sizes. The experimental data used for comparison (black line) are from Glenn Research Center's MGRU3 experiments with the GRC-1 simulant.}\label{fig:comp_size}
\end{minipage}
\end{figure}

In Sec.~\ref{sec:soil_props}, it was pointed out that the DS is 20 times larger than the GRC-1 simulant. In this section, we demonstrate that shrinking the clump size does not affect significantly the simulation accuracy, see Fig.~\ref{fig:comp_size}.

We repeat the test where the \SI{22}{kg} wheel operates at an angular velocity of $\SI{0.8}{rad/s}$ under Earth's gravity, with the terrain made of two new types of DEM elements. 
The blue curve shows the slip--slope relationship when each DS element's size is scaled by a $0.75$ factor to obtain a different digital simulant.  This reduces the apparent size of the smallest type of clumps from \SI{2.5}{mm}  to \SI{1.875}{mm}. \SI{2846244}{} clumps employing \SI{8707887}{} are in the simulation and this \SI{8.4}{s} simulation took approximately \SI{11}{} hours. 
Note that this size change requires re-running the entire sample generation and test bed preparation process before carrying out the slope vs. slip test. Compared to the base red curve, we did not observe a significant change in the slope vs. slip results. 
Conversely, the green curve shows slope vs. slip results when the  DS element size is increased by a factor of $2$, with \SI{299929}{} clumps (\SI{917907}{} sphere components) participating in the simulation. A time of \SI{1.5}{} hours is needed for this \SI{8.4}{s} simulation. 
A rendering of this test is shown in Fig.~\ref{fig:sph_wheel_bigger}. In simulations, larger clumps in this scenario reduced flowability with geometric locking, which led to the wheel slipping less, especially on steeper slopes.

In conclusion, although DS has a particle size that is on average 20 times larger than the particle size in GRC-1, reducing the DS element size does not negatively impact the simulation fidelity. However, if the DS element sizes are increased by a factor of 2 (i.e. simulation resolution is reduced), the accuracy is hindered due to the induced low flowability and geometric locking.

\subsubsection{Sensitivity of slope vs. slip curve with respect to the friction coefficient}

\begin{figure}[htp!]
	\centering
	\captionsetup{justification=centering}
	\includegraphics[width=.9\linewidth]{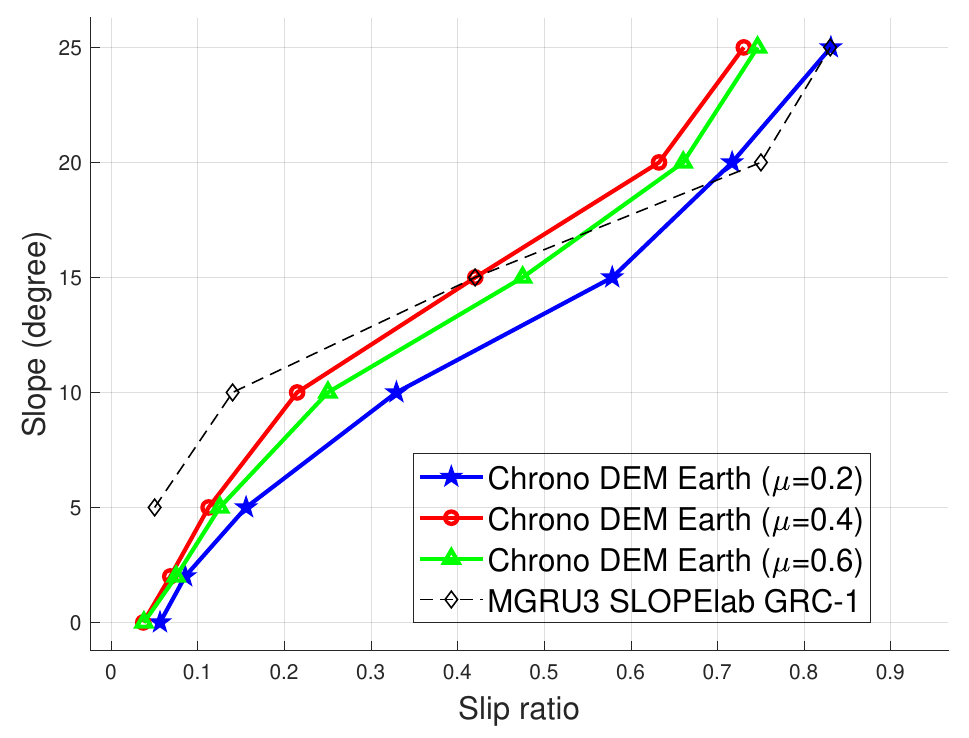}
	\caption{The single-wheel slope--slip test results with different selections of the friction coefficient. The experimental data used for comparison (black line) are from Glenn Research Center's MGRU3 experiments with the GRC-1 simulant.} \label{fig:mu_effect}
\end{figure}

The choice of friction coefficient $\mu=0.4$ is a typical value for silica--silica contact. However, it is well-known that the actual value of the friction coefficient can vary depending on the environment, e.g., humidity and preparation of the facets that come in contact \cite{BowdenFriction,Deng1992ASO}. In this subsection, the focus is on investigating the sensitivity of the slope vs. slip curve with respect to the friction coefficient. To that end, we repeat the test where the \SI{22}{kg} wheel operates at an angular velocity of $\SI{0.8}{rad/s}$ under Earth's gravity (red curve in Fig.\ref{fig:wheel_res_a}), with two different inter-element friction coefficient values. The blue curve in Fig.\ref{fig:mu_effect} shows results when $\mu=0.2$ -- one can observe significantly more slip compared to the base case of $\mu=0.4$. However, in the case where $\mu=0.6$, the slip was comparable to the base case. This suggests that increasing $\mu$ helps prevent the wheel from slipping on the incline, up to a certain value, beyond which the influence of $\mu$ seems to diminish and other factors, e.g., particle locking, start playing an important role.

\subsubsection{Sensitivity of slope vs. slip curve with respect to the clump shape}

The strength of the terrain in mobility tests is expected to change as the clump shapes move away from spheres. In this section, we report simulation results that quantify this change. 
One extreme is to have monodisperse granular material. To test this scenario, we repeat the single-wheel tests in Sec.~\ref{sec:sw} with the radius of the spheres being first \SI{3}{mm} and then \SI{6}{mm}. The wheel has a constant angular velocity of \SI{0.8}{rad/s} and the gravitational acceleration is $g=\SI{9.81}{m/s^2}$. The material properties of the granular material are as in Table~\ref{tab:GRCDS}, except that the friction coefficient $\mu$ is increased from $0.4$ to $0.9$. This change is motivated by difficulties making the spherical elements rest on a steep incline when using $\mu=0.4$. The latter friction coefficient, associated with dry silica, is acceptable for the DS tests since the elements' irregular shapes contribute to the locking of the grains.

\begin{figure}[htp!]
\centering
\begin{minipage}{.4\textwidth}
	\centering
	\captionsetup{justification=centering}
	\includegraphics[width=\linewidth]{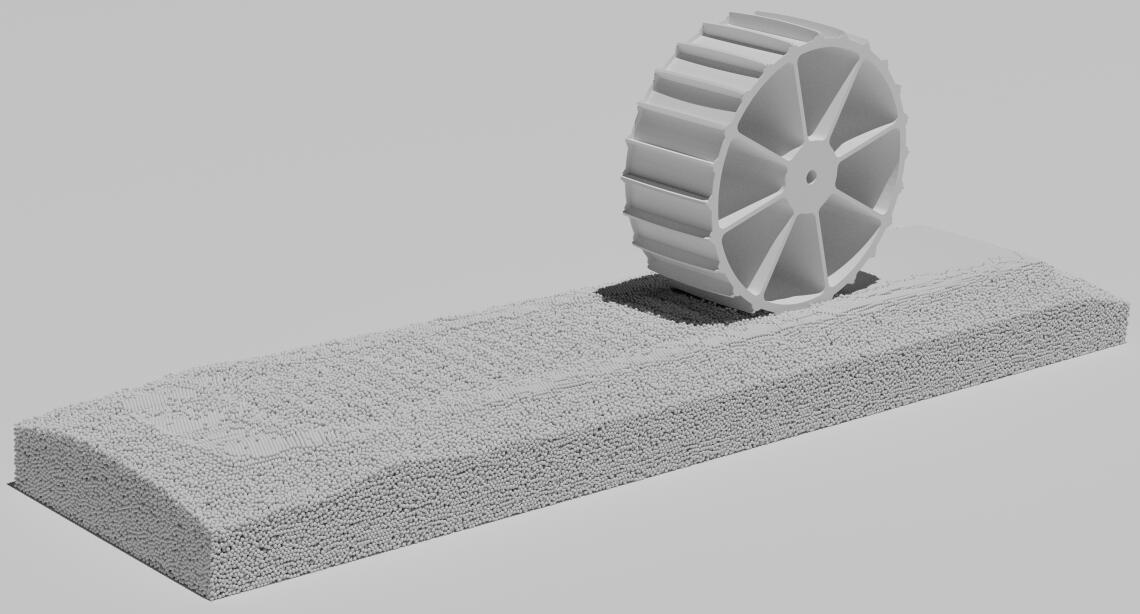}
	\caption{A rendering of the single-wheel test with the granular terrain represented by spherical elements, at $0^\circ$ incline.} \label{fig:sph_wheel_a}
\end{minipage}%
\hspace{.08cm}
\begin{minipage}{.56\textwidth}
	\centering
	\captionsetup{justification=centering}
	\includegraphics[width=\linewidth]{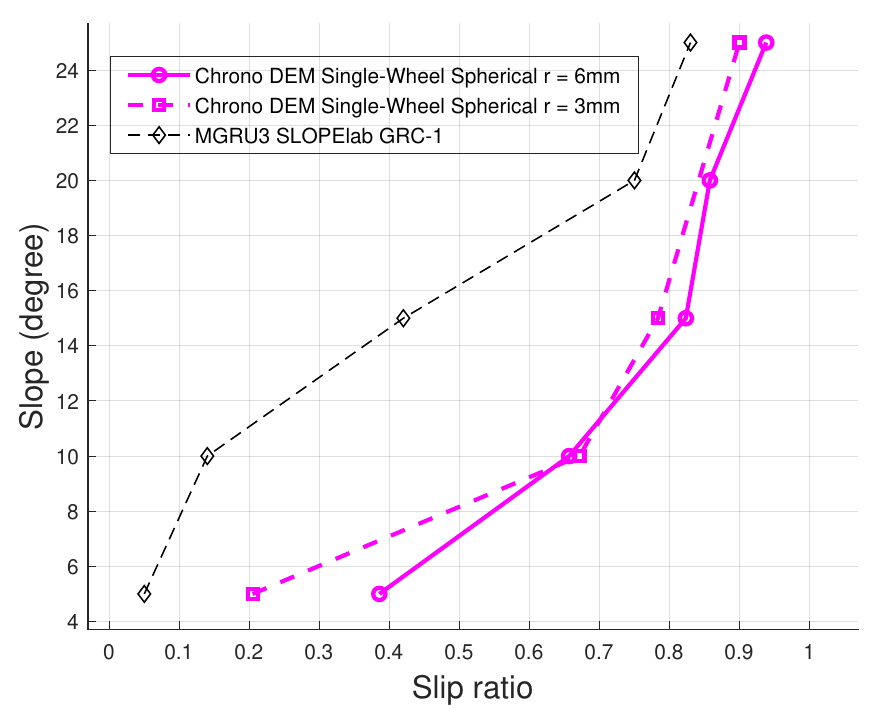}
	\caption{Slip-on-incline test result with spherical terrain elements of radius \SI{6}{mm} and \SI{3}{mm}. The experimental data used for comparison (black line) are from Glenn Research Center's MGRU3 experiments with the GRC-1 simulant.}\label{fig:comp_sph}
\end{minipage}
\end{figure}

The slope vs. slip relationship is summarized in Fig.~\ref{fig:comp_sph}. If the spherical elements are fine enough (\SI{3}{mm} in this test, the dashed magenta line), the sphere-based DEM appears to show a somewhat reasonable slip ratio in the flat terrain tests. A rendering is shown in Fig.~\ref{fig:sph_wheel_a}. However, even for mobility tests on a gentle incline, the monodisperse terrain shears drastically more than the DS and the real-world GRC-1 simulant. The wheel is essentially spinning in place at a $15^\circ$ incline. At a $25^\circ$ incline, the elements immediately flow down the incline upon making contact with the wheel as previously shown in Sec.~\ref{sec:intro}, Fig.~\ref{fig:SPHat25}.

\begin{figure}[htp!]
\centering
\begin{minipage}{.4\textwidth}
	\centering
	\captionsetup{justification=centering}
	\includegraphics[width=\linewidth]{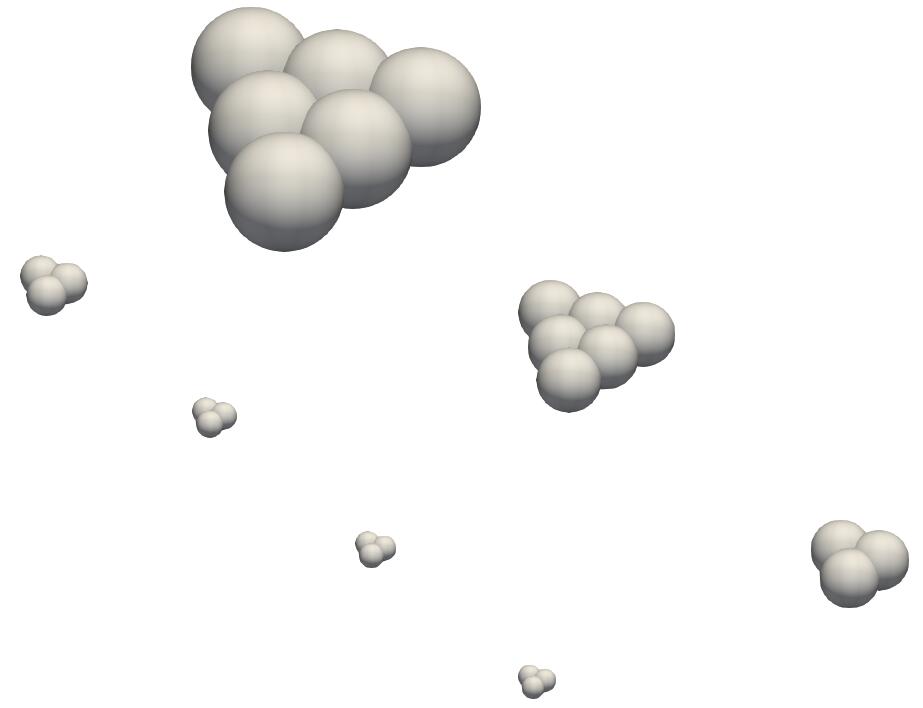}
	\caption{The clump shapes after enlarging their aspect ratio.} \label{fig:long_seven_types}
\end{minipage}%
\hspace{.08cm}
\begin{minipage}{.56\textwidth}
	\centering
	\captionsetup{justification=centering}
	\includegraphics[width=\linewidth]{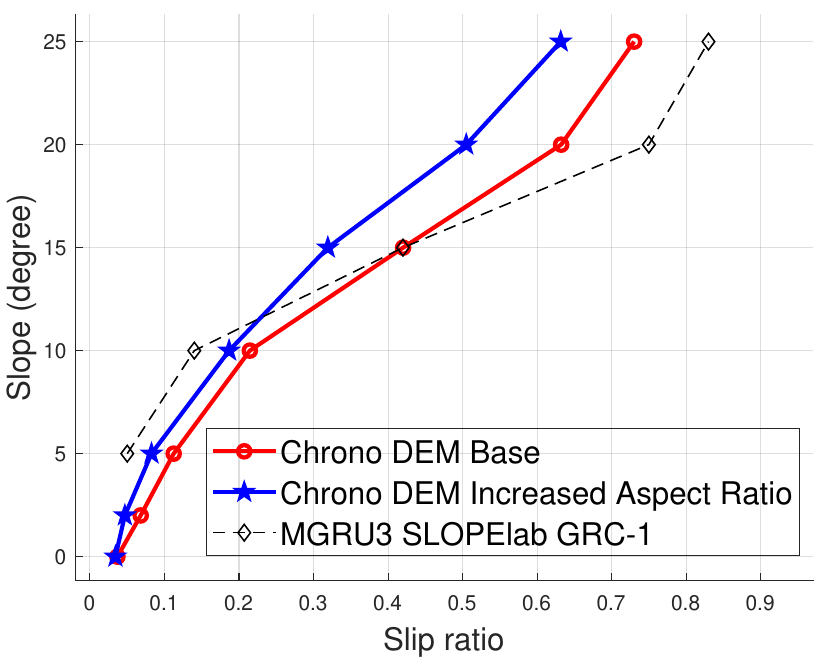}
	\caption{Slip-on-incline test result with the clumps that have larger aspect ratios, compared against the base clump shape. The experimental data used for comparison (black line) are from Glenn Research Center's MGRU3 experiments with the GRC-1 simulant.}\label{fig:comp_elong}
\end{minipage}
\end{figure}

We also present an opposite investigation where the aspect ratios of the clumps are enlarged by expanding the clumps without changing their thickness.
This is accomplished by moving each component sphere slightly away from the geometric center of the clump while maintaining the planar overall shape of the clump, see Fig.~\ref{fig:long_seven_types} for new shapes. Quantitatively, keeping other conditions unchanged, the seven clump templates' aspect ratio was increased by about $25\%$. After creating a new DEM terrain using the new clumps, the single-wheel test is repeated with a wheel angular velocity of \SI{0.8}{rad/s} and gravitational pull of $g=\SI{9.81}{m/s^2}$. We observe in Fig.~\ref{fig:comp_elong} that the slip, shown with the blue curve, is reduced compared to the base red curve, and this effect is more pronounced on steeper slopes.

Moving from spherical to clumped DEM elements has a significant impact on terrain strength in mobility simulations. The geometric locking effect that DS enables on incline tests cannot be replicated using monodisperse terrain. We note the existence of studies on reproducing the locking effect in monodisperse material with modified material properties such as the rolling resistance~\cite{XIE2019207}. Its applicability in rover mobility simulations is unclear, but even if feasible, it necessitates extra calibration stages that require field test data for each terrain configuration of interest. By using clumps, the DS does not require field test data but rather a good approximation of the elements' shapes.

Finally, note that dilating the clump at constant thickness also reduces the slope associated with a given slip, albeit not as significantly as noted for monodisperse material. The findings also suggest that if high fidelity is needed, one should pay attention to the shape of the elements. This can be non-trivial, particularly if the granular material does not have a known shape distribution, an aspect that can be addressed through a shape calibration process that falls outside the scope of this contribution.

\subsection{Full-rover slip test on incline} \label{sec:full_rover}

\begin{figure}[H]
	\centering
	\captionsetup{justification=centering}
	\subfloat[A rendering of the VIPER rover used in the simulation.\label{fig:full_a}]
	{
		\includegraphics[width=0.3\linewidth]{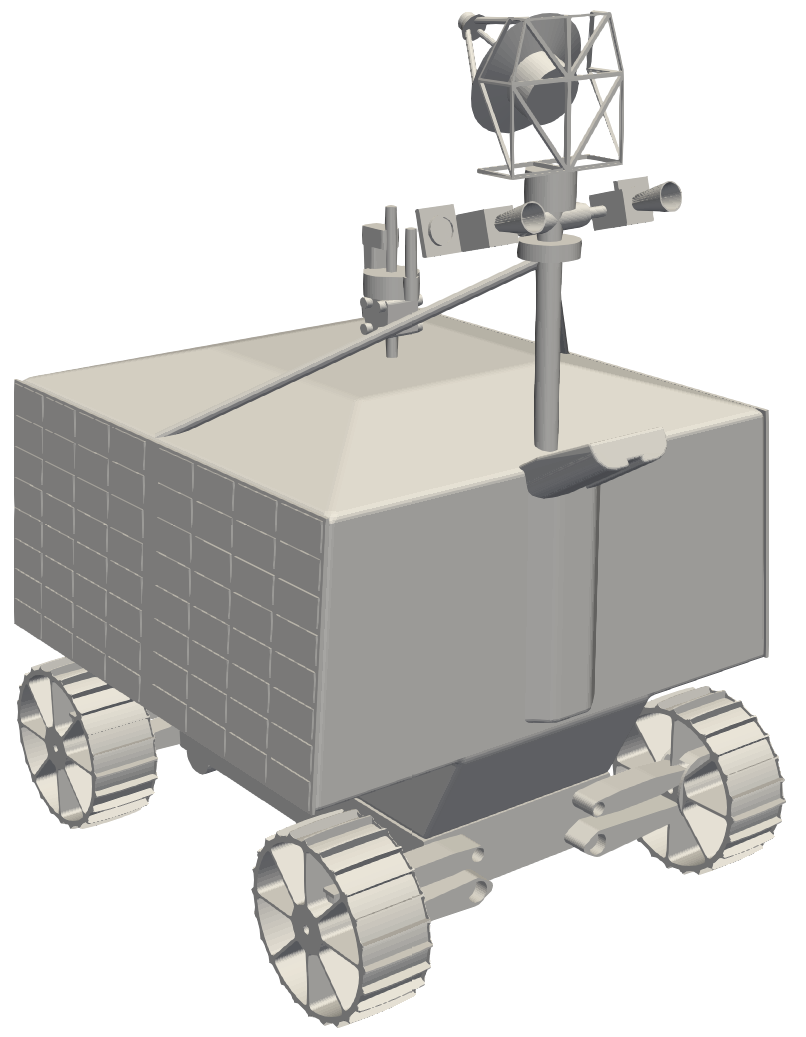}
	}
%
	\hspace{.1cm}
	\subfloat[A rendering of the VIPER rover operating on a $20^\circ$ incline.\label{fig:full_d}]
	{
		\includegraphics[width=0.6\linewidth]{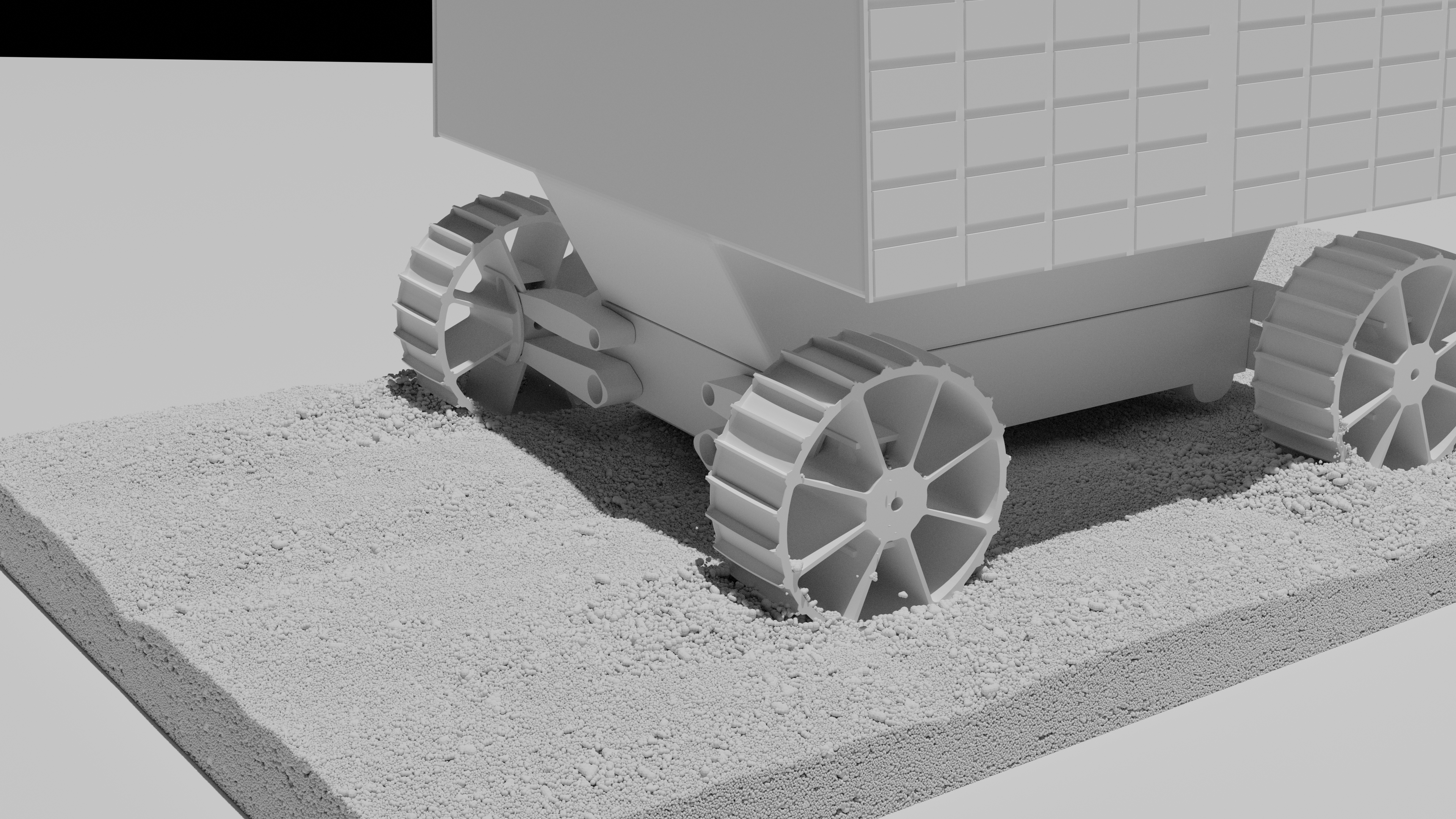}
	}
	\caption{The full-rover test using the DS.} 
\end{figure}

This section discusses a slope vs. slip test carried out for a full rover. This calls for a co-simulation between a multi-body system (the rover) and a DEM system (the terrain). For the multi-body system, since the MGRU3 CAD model was inaccessible, we instead used a similar VIPER rover model publicly available in the latest  Chrono distribution~\cite{chronoOverview2016}. Shown in Fig.~\ref{fig:full_a}, the rover is roughly \SI{2.1}{m} in length, \SI{1.5}{m} in width, and \SI{1.4}{m} in height (excluding the antenna). Four wheels are connected to the rover body through revolute joints. The wheel geometry is from MGRU3, the same as that in Sec.~\ref{sec:sw}. The rover moves around by prescribing all its four wheels a \SI{0.8}{rad/s} angular velocity on inclines of 0, 5, 10, 15, 20, and 25$^\circ$. A rendering of a test scene is provided in Fig.~\ref{fig:full_d}.

\begin{figure}[htp!]
	\centering
	\captionsetup{justification=centering}
	\includegraphics[width=.9\linewidth]{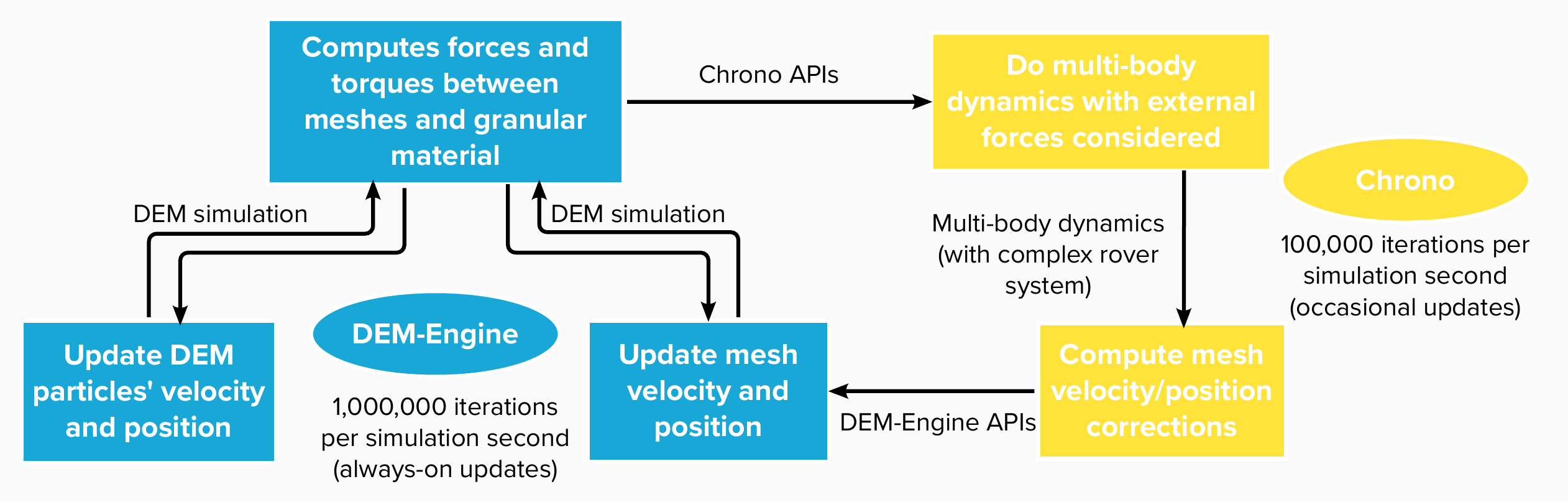}
	\caption{The co-simulation workflow between the multi-body system simulated by Chrono and the DEM system simulated by DEM-Engine.} \label{fig:cosim_workflow}
\end{figure}

The co-simulation setup is shown in Fig.~\ref{fig:cosim_workflow}.  The DEM simulator handles the evolution of the granular terrain, while Chrono deals with the rover dynamics. The two simulators are bridged through the meshes that represent the wheels. The DEM simulator calculates the force exerted by the terrain on the wheel mesh. The force information is used when the Chrono numerical integrator propagates forward in time the evolution of the meshes. In turn, the new position of the wheels will become a set of updated boundary conditions for the granular material.  
The rover's mobility is also influenced by forces that crop up in the chassis and suspension that are independent of the motion of the granular terrain.
In the current Chrono co-simulation framework, we advance the multi-body rover system in Chrono by one time step for every ten DEM time steps. 

\begin{figure}[htp!]
	\centering
	\captionsetup{justification=centering}
	\includegraphics[width=.9\linewidth]{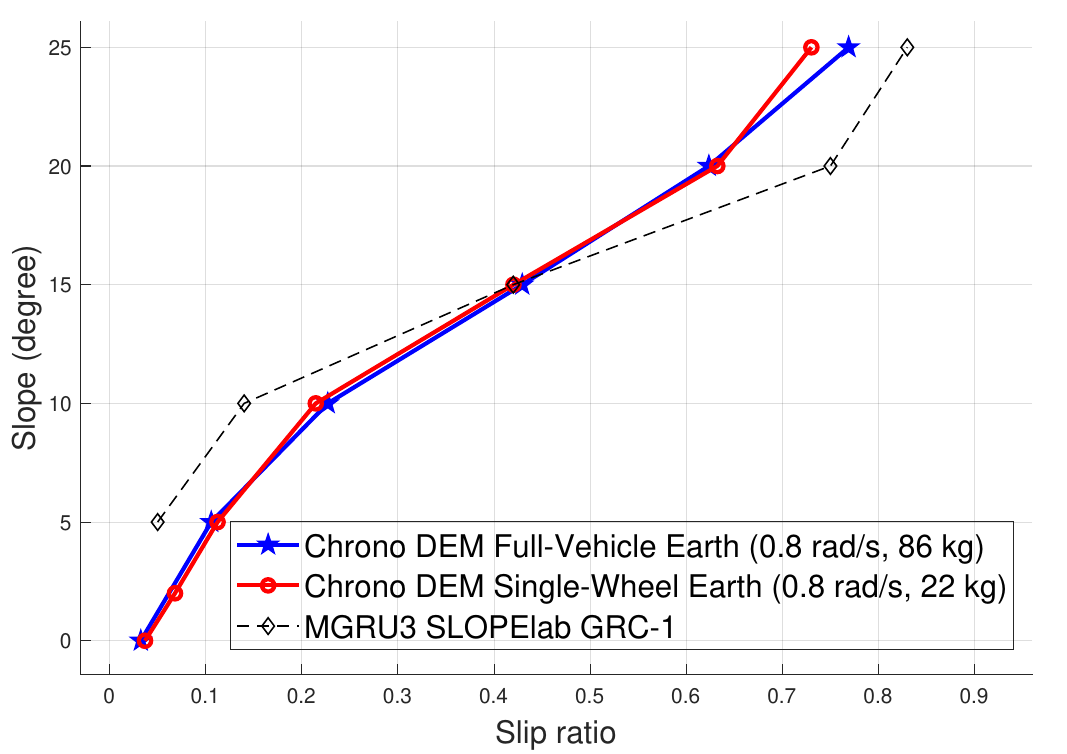}
	\caption{VIPER rover slip-on-incline test result, compared against the single-wheel test done in the same conditions. The experimental data used for comparison (black line) are from Glenn Research Center's MGRU3 experiments with the GRC-1 simulant.} \label{fig:full_res}
\end{figure}

Note that the full-rover slip data shown in Fig.~\ref{fig:full_res} displays no notable difference compared to the single-wheel counterpart. This suggests that the more expeditious single-wheel tests are likely sufficient to gain insights into the rover's mobility attributes. The slip ratio increases relatively slowly with the slope angle in the interval between $0^\circ$ and $10^\circ$. Past $10^\circ$, this rate of increase escalates, and the rover almost fails to climb on a $25^\circ$ incline. Finally, the DS for this test uses \SI{11336638}{} DEM elements that employ \SI{34691952}{} component spheres.
On two NVIDIA A100 GPUs, a \SI{15}{s} long simulation requires approximately \SI{109}{} hours of run time.

\section{Summary and future work} \label{sec:summary}
We report results obtained with a new GPU-accelerated DEM solver that uses a clump-based representation of the discrete elements. The new DEM simulator is showcased in conjunction with a granular material similar to the GRC-1 simulant and which is modeled using elements of nontrivial shapes. Given that particle sizes in GRC-1 go down to micron-level, the DS used herein qualitatively embraces the element size distribution of GRC-1 but increases all element sizes by a factor of 20. Despite this size augmentation, the DS is shown to behave similarly to GRC-1 in several tests, i.e., angle of repose, cone penetration, drawbar pull, and single-wheel and full-rover incline slip vs. slope tests. 
The DEM simulations scale up to tens of millions of elements that have nontrivial shapes obtained by clumping together spheres of various radii. The DEM simulator works in the Chrono framework and leverages two GPUs simultaneously. The simulator and the code for all simulations discussed are open-source and available at \cite{RuochunDEMERepo}.

A salient observation is that the simulation methodology advanced by the DEM simulator is physics-based. One only needs to specify the parameters $E$, $\nu$, $\mu$, and CoR to be able to carry out the simulation. Moreover, these parameters can be specified on a per-sphere basis rather than on a per-element basis, which opens the door to simulating complex discrete elements that involve particles of different shapes and different materials. To the best of our knowledge, our simulator is unique in this regard. Finally, particle breakage is supported yet it will be discussed in a follow up contribution.

There are three directions in which this effort will be continued. First, we plan to  generate models associated with other extraterrestrial simulants, e.g., GRC-3 and Mojave Mars simulant \cite{PETERS2008470}, and then assess their terramechanics attributes following the workflow established in this contribution. In this context, when moving to full rover analysis or heavy wheels, we will have to also increase in simulation the depth of the bed of granular material. For the \SI{22}{kg} single wheel tests, the thickness of the bed was \SI{15}{cm}. For heavy implements, this value will have to be increased to avoid boundary effects noted in practice. Second, we are carrying out a systematic study to gauge how various wheel design attributes, e.g., wheel shape/size/width, and grouser count/height/width/shape, influence the mobility of a rover operating on lunar terrain. The focus is on how a wheel design changes rover attributes such as energy consumption, steering, slip initiation, and climbing ability. Lastly, we are interested in understanding how the high-fidelity DEM simulation results can be used to design data-driven, expeditious wheel-terrain interaction methods.

\section*{Declarations} 
\subsection*{Funding} This work has been partially supported by NSF projects OAC2209791 and CISE1835674, US Army Research Office project W911NF1910431, and NASA Small Business Technology Transfer (STTR) Sequential Contract \#80NSSC20C0252. 
\subsection*{Competing interests} The authors have no competing interests to declare that are relevant to the content of this article.
\subsection*{Other contributor} The authors would like to thank Colin Creager of NASA Glenn for his insights and feedback on this manuscript.
\subsection*{Code availability} The code used to reproduce the results in this work is available in a GitHub public repository \cite{DEMERepoGRCBranch}. The DEM simulator used in this work is also available in a GitHub public repository \cite{RuochunDEMERepo} for unrestricted use and distribution.

\bibliography{refs}

\end{document}